\documentclass[]{aa}  

\usepackage{graphicx}
\usepackage{txfonts}
\usepackage{hyperref}
\pdfoutput=1

\usepackage{xspace}
\usepackage{color}
\usepackage{booktabs}

\usepackage{environ}
\NewEnviron{splitequation}{%
    \begin{equation}\begin{split}
        \BODY
    \end{split}\end{equation}}

\newcommand{\software}[1]{\textsc{#1}}

\newcommand{\bb}[1]{\left[ #1 \right]}
\newcommand{\Var}{\mathrm{Var}}

\newcommand{\diff}{{\rm d}}

\begin{document}

   \title{KiDS-1000 Cosmology: constraints beyond flat $\Lambda$CDM}

   \authorrunning{Tr\"oster \& the KiDS Collaboration et al.}

   \author{Tilman Tr\"oster\inst{1}\thanks{Email: ttr@roe.ac.uk} 
   \and Marika Asgari\inst{1} 
   \and Chris Blake\inst{2}
   \and Matteo Cataneo\inst{1}
   \and Catherine Heymans \inst{1,3}
   \and Hendrik Hildebrandt\inst{3}
   \and Benjamin Joachimi\inst{4}
   \and Chieh-An Lin\inst{1}
   \and Ariel~G.~S\'anchez\inst{5}
   \and Angus~H.~Wright\inst{3}
   \and Maciej Bilicki\inst{6}
   \and Benjamin Bose\inst{7}
   \and Martin Crocce\inst{8,9}
   \and Andrej Dvornik\inst{3}
   \and Thomas Erben\inst{10}
   \and Benjamin Giblin\inst{1}
   \and Karl Glazebrook\inst{2}
   \and Henk Hoekstra\inst{11}
   \and Shahab Joudaki\inst{12}
   \and Arun Kannawadi\inst{13}
   \and Fabian K\"ohlinger\inst{3}
   \and Konrad Kuijken\inst{11}
   \and Chris Lidman\inst{14,15}
   \and Lucas Lombriser\inst{7}
   \and Alexander Mead\inst{16}
   \and David Parkinson\inst{17}
   \and HuanYuan Shan\inst{18,19}
   \and Christian Wolf \inst{14,15}
   \and Qianli Xia\inst{1}
          }
\institute{Institute for Astronomy, University of Edinburgh, Royal Observatory, Blackford Hill, Edinburgh, EH9 3HJ, UK 
   \and 
   Centre for Astrophysics \& Supercomputing, Swinburne University of Technology, P.O. Box 218, Hawthorn, VIC 3122, Australia
   \and
   Ruhr-Universit{\"a}t Bochum, Astronomisches Institut, German Centre for Cosmological Lensing (GCCL), Universit{\"a}tsstr.  150, 44801, Bochum, Germany
   \and
   Department of Physics and Astronomy, University College London, Gower Street, London WC1E 6BT, UK
   \and
   Max-Planck-Institut f\"ur extraterrestrische Physik, Postfach 1312, Giessenbachstrasse 1, D-85741 Garching, Germany
  \and
  Center for Theoretical Physics, Polish Academy of Sciences, al. Lotnik\'{o}w 32/46, 02-668, Warsaw, Poland
  \and
  Département de Physique Théorique, Université de Genève, 24 quai Ernest Ansermet, 1211 Genève 4, Switzerland
  \and 
  Institute of Space Sciences (ICE, CSIC), Campus UAB, Carrer de Can Magrans, s/n,  E-08193 Barcelona, Spain 
  \and
  Institut d'Estudis Espacials de Catalunya (IEEC),  Carrer Gran Capita 2, E-08034, Barcelona, Spain
  \and
  Argelander-Institut f\"ur Astronomie, Universit\"at Bonn, Auf dem H\"ugel 71, D-53121 Bonn, Germany
  \and
  Leiden Observatory, Leiden University, Niels Bohrweg 2, 2333 CA Leiden, the Netherlands
  \and
  Department of Physics, University of Oxford, Denys Wilkinson Building, Keble Road, Oxford OX1 3RH, UK
  \and
  Department of Astrophysical Sciences, Princeton University, 4 Ivy Lane, Princeton, NJ 08544, USA
  \and
  Centre for Gravitational Astrophysics, College of Science, Australian National University, ACT 2601, Australia
  \and
  Research School of Astronomy and Astrophysics, Australian National University, Canberra ACT 2600, Australia
  \and
  Institut de Ciències del Cosmos, Universitat de Barcelona, Martí Franquès 1, E08028 Barcelona, Spain
  \and
  Korea Astronomy and Space Science Institute, 776 Daedeokdae-ro, Yuseong-gu, Daejeon 34055, Republic of Korea
    \and
  Shanghai Astronomical Observatory (SHAO), Nandan Road 80, Shanghai 200030, China
  \and
  University of Chinese Academy of Sciences, Beijing 100049, China
   }

   \date{}

\abstract{
We present constraints on extensions to the standard cosmological model of a spatially flat Universe governed by general relativity, a cosmological constant ($\Lambda$), and cold dark matter (CDM) by varying the spatial curvature $\Omega_K$, the sum of the neutrino masses $\sum m_\nu$, the dark energy equation of state parameter $w$, and the Hu-Sawicki $f(R)$ gravity $f_{R0}$ parameter.
With the combined $3\times2$pt measurements of cosmic shear from the Kilo-Degree Survey (KiDS-1000), galaxy clustering from the Baryon Oscillation Spectroscopic Survey (BOSS), and galaxy-galaxy lensing from the overlap between KiDS-1000, BOSS, and the spectroscopic 2-degree Field Lensing Survey (2dFLenS), we find results that are fully consistent with a flat $\Lambda$CDM model with $\Omega_K=0.011^{+0.054}_{-0.057}$, $\sum m_\nu<1.76\,{\rm eV}$ (95\% CL), and $w=-0.99^{+0.11}_{-0.13}$. 
The $f_{R0}$ parameter is unconstrained in our fully non-linear $f(R)$ cosmic shear analysis. 
Considering three different model selection criteria, we find no clear preference for either the fiducial flat $\Lambda$CDM model or any of the considered extensions. 
In addition to extensions to the flat $\Lambda$CDM parameter space, we also explore restrictions to common subsets of the flat $\Lambda$CDM parameter space by fixing the amplitude of the primordial power spectrum to the \textit{Planck} best-fit value, as well as adding external data from supernovae and lensing of the cosmic microwave background (CMB).
Neither the beyond-$\Lambda$CDM models nor the imposed restrictions explored in this analysis are able to resolve the $\sim3\,\sigma$ tension in $S_8$ between the $3\times2$pt constraints and the \textit{Planck} temperature and polarisation data, with the exception of $w$CDM, where the $S_8$ tension is resolved. 
The tension in the $w$CDM case persists, however, when considering the joint $S_8$--$w$ parameter space.   
The joint flat $\Lambda$CDM CMB lensing and $3\times2$pt analysis is found to yield tight constraints on $\Omega_{\rm m}=0.307^{+0.008}_{-0.013}$, $\sigma_8=0.769^{+0.022}_{-0.010}$, and $S_8=0.779^{+0.013}_{-0.013}$.
}


   \maketitle
%

\section{Introduction}
A wide range of cosmological observations support a theoretical model for the Universe comprised of cold dark matter (CDM) and a cosmological constant ($\Lambda$), with baryons very much in the minority. 
These components are connected through a spatially flat gravitational framework within general relativity. 
This flat $\Lambda$CDM model can independently describe the temperature fluctuations in the cosmic microwave background \citep[CMB,][]{Planck2020-Cosmology}, the baryon acoustic oscillation and redshift-space distortions in the clustering of galaxies \citep[BAO and RSD,][]{Alam2017,eBOSS2020}, the accelerating expansion rate seen in the distance-redshift relation of Type Ia supernovae \citep[SNe,][]{Scolnic2018}, the present-day expansion rate as measured using a distance ladder calibrated through Cepheid variables \citep{Riess2019} or strongly lensed quasars \citep{Wong2020}, and the weak gravitational lensing of background light by foreground large-scales structures \citep{Troxel2018, HSC-xi, Asgari2021-CS, Planck2020-CMBlensing}.  

The flat $\Lambda$CDM model is highly successful in describing these observables independently, but differences arise in the precise values of some cosmological components when analysing certain probes in combination. 
In comparison to values predicted from the best-fitting flat $\Lambda$CDM model to observations of the CMB \citep{Planck2020-Cosmology}, \citet{Riess2019} and \citet{Wong2020} report $\sim 4$--$5\,\sigma$ differences in direct local measurements of the Hubble parameter $H_0$, with other measurements, such as the inverse distance ladder \citep{eBOSS2020} or the tip of the red giant branch \citep{Freedman2020}, lying in between. 
\citet{Asgari2021-CS} report $\sim 3\,\sigma$ differences in $S_8 = \sigma_8 \sqrt{\Omega_{\rm m}/0.3}$, which is a direct measure of the clustering and density of large-scale structures, following the trend to lower $S_8$ values seen in other weak gravitational lensing surveys \citep[e.g.][]{Heymans2013, Troxel2018, HSC-Cl}. 
Provided that all sources of systematic uncertainty have been accounted for in each analysis, the tensions reported between early and late-time probes of the Universe can be considered as potential evidence for the existence of additional components in our cosmological model, beyond flat $\Lambda$CDM. 

Such extensions have been considered before \citep[e.g.][]{Planck2016-fR,Joudaki2017-KiDS450-ext,DES-extended,Planck2020-Cosmology,eBOSS2020,Dhawan2020}, with no strong evidence for a Universe that deviates from flat $\Lambda$CDM with a minimal neutrino mass. 
While the combination of CMB and large-scale structure data rules out strong deviations from a flat $\Lambda$CDM model, the constraints from just the early or late-time Universe are much weaker, with \textit{Planck} data favouring a closed Universe \citep[e.g.][]{Planck2020-Cosmology,OOba2018,Park2019,Handley2019b,DiValentino2020} but see also \citet{Efstathiou2020} for a different view.

Here we explore extensions to the flat $\Lambda$CDM model independently of CMB temperature and polarisation data, presenting constraints on the cosmological parameters that describe four separate additions. 
We allow for non-zero curvature ($o\Lambda$CDM), include uncertainty in the sum of the neutrino masses ($\nu\Lambda$CDM), replace the cosmological constant with an evolving dark energy component ($w$CDM), and explore modifications to standard gravity using the \citet{Hu2007} $f(R)$-gravity model, where the gravitational force is enhanced in low-density regions. 

To confront this range of models, we compare CMB temperature and polarisation observations\footnote{Unless otherwise specified, `\textit{Planck} data' shall refer to the primary anisotropy data of the \citet{Planck2020-Cosmology} \mbox{TTTEEE+lowE} likelihood.}  from \citet{Planck2020-Cosmology} to different combinations of late Universe probes. 
We analyse the weak gravitational lensing of galaxies, imaged by the fourth data release of the Kilo-Degree Survey \citep[KiDS-1000,][]{Kuijken2019}, the gravitational lensing of the CMB \citep{Planck2020-CMBlensing}, Type Ia SNe \citep{Scolnic2018}, and galaxy clustering observations from the twelfth data release of the Baryon Oscillation Spectroscopic Survey \citep[][]{Alam2017}. 

In Sect.~\ref{sec:datamethodology}, we summarise the cosmological observations that we analyse in this paper, as well as the methodology. 
We introduce the $\Lambda$CDM extensions that we adopt in Sect.~\ref{sec:models} and present our model constraints in Sect.~\ref{sec:results}. 
We conclude our analysis in Sect.~\ref{sec:conclusions}. 
In the appendices we demonstrate that our constraints on $S_8$ are insensitive to two potential sources of systematic error in our analysis. 
In Appendix~\ref{app:HMCODE} we compare parameter constraints using two different models to account for our uncertainty on how baryon feedback impacts the shape of the non-linear matter power spectrum. 
In Appendix~\ref{app:nscuts} we exclude large-scale information from the galaxy clustering observable and introduce informative priors on the tilt of the primordial power spectrum, $n_{\rm s}$. 

\section{Data and methodology}
\label{sec:datamethodology}
The data and methodology, unless mentioned otherwise, match those presented by \citet{Heymans2021}. 
Here we summarise the salient points and refer the reader to \citet{Joachimi2021} for details about the methodology, \citet{Asgari2021-CS} for the cosmic shear analysis, and \citet{Heymans2021} for an in-depth description of the multi-probe analysis of KiDS, BOSS, and 2dFLenS.

\subsection{KiDS, BOSS, and 2dFLenS data}
The fourth data release of the Kilo-Degree Survey images $1006\,\mathrm{deg}^{2}$ in nine bands, spanning the optical to the near-infrared \citep{Kuijken2019}.
The survey strategy is optimised for weak lensing observations with accuracy and precision in the shear and redshift estimates aided by high-resolution deep imaging in the $r$-band, a camera with a smoothly varying and low-ellipticity point-spread function, complete matched-depth observations across the full wavelength range \citep{Wright2019}, and auxiliary imaging of deep spectroscopic calibration fields. 
\citet{Giblin2021} present the KiDS-1000 weak lensing shear catalogue, along with a series of null tests to quantify any systematic signals associated with the instrument, verifying that they do not introduce any bias in a cosmological analysis. 
\citet{Hildebrandt2020-CZ} present the KiDS-1000 photometric redshift estimates for the `gold' galaxy sample, selected to ensure complete representation in the spectroscopic calibration sample \citep{Wright2020}. 
The resulting redshift distributions are validated using measurements of galaxy clustering between spectroscopic and photometric samples \citep{vandenbusch2020,Hildebrandt2020-CZ}.

The Baryon Oscillation Spectroscopic Survey \citep[BOSS,][]{Dawson2013} of a sample of 1.2 million luminous red galaxies (LRGs) over an effective area of $9329\,\mathrm{deg}^{2}$ provides the optimal data set to observe large-scale galaxy clustering at high signal-to-noise out to redshift $z<0.75$. 
\citet{Alam2017} present a compilation of different statistical analyses of the baryon acoustic oscillation peak and the redshift-space distortions of the twelfth data release (DR12) of the BOSS sample. 
Combined with CMB observations from \citet{Planck2016}, \citet{Alam2017} set constraints on $o\Lambda$CDM, $w$CDM and $\nu\Lambda$CDM cosmological models, with the joint data set showing no preference for extending the cosmological model beyond flat $\Lambda$CDM. 
The same conclusion is drawn, with improved precision, in the recent \citet{eBOSS2020} galaxy clustering analysis. 
This extended-BOSS survey includes galaxy and quasar samples out to $z <2.2$, and Lyman-$\alpha$ forest observations between $2<z<3.5$. 

The `galaxy-galaxy lensing' (GGL) of background KiDS galaxies by foreground LRGs is measured on the overlapping areas of KiDS with BOSS DR12 and the 2-degree Field Lensing Survey \citep[2dFLenS,][]{Blake2016}. 
2dFLenS covers $731\,\mathrm{deg}^{2}$, with spectroscopic redshifts for $70\,000$ galaxies out to $z<0.9$ and was designed to target areas already mapped by weak lensing surveys to facilitate `same-sky' lensing-clustering analyses \citep{johnson/etal:2017,amon/etal:2018,Joudaki2018, Blake2020}.

Cosmological constraints on the parameters of the flat $\Lambda$CDM analysis of KiDS-1000 are presented in \citet{Asgari2021-CS} and \citet{Heymans2021}. 
\citet{Asgari2021-CS} analyse the observed evolution of weak lensing by large-scale structures, referred to as cosmic shear, in five redshift bins, using a range of different two-point statistics. 
\citet{Heymans2021} combine these cosmic shear measurements with BOSS DR12 galaxy clustering observations from \citet{Sanchez2017} and GGL observations of KiDS-1000 galaxies by LRGs from BOSS and 2dFLenS. 
The combination of these three two-point large-scale structure probes is often referred to as `$3\times2$pt', with the methodology described and validated using a large suite of mock survey catalogues in \citet{Joachimi2021}. 

We choose angular power spectrum estimates for our cosmic shear and GGL summary statistics, following \citet{Heymans2021}.
Specifically, we use the `band power' estimator, a linear transformation of the real-space two-point correlation functions \citep{Schneider2002}, and estimate the angular shear and GGL power spectra in eight logarithmically spaced bands between $\ell = 100$ and $\ell = 1500$, for five tomographic redshift bins between $z=0.1$ and $z=1.2$, and the two spectroscopic lens bins $z\in (0.2,0.5]$ and $z\in(0.5,0.75]$.
We discard GGL measurements at small scales and where there is overlap between the source and lens bins due to limitations in our modelling of non-linear galaxy bias and intrinsic alignment.

Our galaxy clustering measurements are adopted from \citet{Sanchez2017}, who analyse the clustering of BOSS galaxies using the anisotropic galaxy correlation function divided into `wedges'. 
We use the two non-overlapping redshift bins of the combined galaxy sample of \citet{Alam2017}, including galaxy separations between $20\,h^{-1}{\rm Mpc}$ and $160\,h^{-1}{\rm Mpc}$. 
In a re-analysis of this data set, \citet{Troester2020} demonstrate that constraints on the flat $\Lambda$CDM model from BOSS clustering alone are fully consistent with \textit{Planck}, but have a preference for lower values of the clustering parameter $S_8$.  
This result is confirmed in two independent BOSS-only re-analyses of the \citet{Beutler2017} Fourier-space BOSS clustering measurements \citep{Ivanov2020, dAmico2020}. 
It is therefore relevant to combine BOSS galaxy clustering constraints with cosmological probes alternative to the CMB, to explore joint constraints on extensions to the flat $\Lambda$CDM model.

\subsection{Likelihood and inference setup}
Our inference pipeline is based on a modified version of \software{CosmoSIS}\footnote{\url{https://bitbucket.org/joezuntz/cosmosis}} \citep{Zuntz2015}, which we call \software{kcap}\footnote{KiDS cosmology analysis pipeline, made public upon acceptance of this paper.}. 
Parameter sampling is performed using \software{MultiNest} \citep{Feroz2008,Feroz2009,Feroz2013}, using 500 or 1000 live points, and an efficiency parameter of 0.3. 
The sampled parameters and priors are summarised in Table~\ref{tab:priors}. 
We vary 12 parameters in our fiducial cosmic shear analysis, 13 parameters for the galaxy clustering analysis, and 20 parameters in our $3\times2$pt analysis.

The linear matter power spectrum and background quantities are calculated using \software{CAMB}\footnote{\url{https://github.com/cmbant/CAMB}} \citep{Lewis2000}, with the non-linear matter power spectrum modelled using \software{hmcode} \citep{Mead2016}. 
The reaction of the non-linear matter power spectrum in the presence of $f(R)$ gravity is modelled using \software{ReACT} \citep{Bose2020}.
The clustering of galaxies uses the same renormalised perturbation theory model employed in \citet{Sanchez2017}, while the non-linear bias for GGL uses the interpolation scheme described in \citet{Joachimi2021, Heymans2021}. 

The covariance of the cosmic shear and GGL data is computed based on the analytical model described in \citet{Joachimi2021}. 
The galaxy clustering covariance is estimated from 2048 mock data realisations \citep[][]{Kitaura2016}, accounting for effect of noise in the covariance on the bias in the inverse Wishart distribution \citep{Kaufman1967, Hartlap2007}. 
As the cross-covariance between our lensing measurements (cosmic shear and GGL) and galaxy clustering is negligible \citep{Joachimi2021}, we treat the lensing and galaxy clustering data vectors as independent.

The maximum of the posterior (MAP) is estimated using the optimisation algorithm of \citet{Nelder1965}, using the 18 samples from the posterior with the highest posterior values as starting points. 
For likelihoods that include the galaxy clustering likelihood, we quote the weighted median of the different MAP runs as the location of the MAP, since numerical noise in the likelihood surface causes poor convergence of the posterior optimisation algorithm \citep{Heymans2021}.

\subsection{Model selection}
\label{sec:moddelselection}
As we consider different models to describe our data, we wish to quantify which of these models describe the data best. 
To this end we make use of three different model selection criteria. 
The individual criteria differ in their dependence on point estimates, priors, and model dimensionalities. 
Considering a range of model selection criteria should therefore lead to a more robust quantification of whether the data prefer one model over another.

The first criterion is the deviance information criterion (DIC, \citealt{Spiegelhalter2002}, for applications in astronomy and cosmology see, for example, \citealt{Kunz2006}, \citealt{Liddle2007}, and \citealt{Trotta2008}):
\begin{equation}
    \label{equ:dicdef}
    \text{DIC} = -2\ln\mathcal{L}(\vec\theta_p) + 2 p_{\rm DIC}\,,\quad p_{\rm DIC} = 2\ln\mathcal{L}(\vec\theta_p) - 2\langle\ln\mathcal{L}\rangle_P \,.
\end{equation}
The first term is given by $-2$ times the logarithm of the likelihood $\mathcal{L}(\vec\theta) = P(\vec d|\vec\theta, M)$ at some point in parameter space $\vec\theta_p$ and encapsulates how well the model fits the data. 
Common choices for $\vec\theta_p$ are the mean, maximum of the posterior, or maximum of the likelihood.
Here we choose $\vec\theta_p$ to be the maximum of the posterior (MAP). 
The second term in Eq.~\eqref{equ:dicdef} is a measure of the model complexity, where the angled brackets denote the average with respect to the posterior $P(\vec\theta |\vec d, M)$. 
When comparing models, those with a lower DIC are preferred.

The second criterion we employ is the Watanabe-Akaike information criterion \citep[WAIC, also known as widely applicable information criterion,][]{Watanabe2010}, a Bayesian generalisation of the DIC, as it does not depend on point estimates and has other, desirable properties \citep{Gelman2014,Vehtari2017}. 
The WAIC is given by
\begin{equation}
    \label{equ:waicdef}
    \text{WAIC} = -2\ln\langle\mathcal{L}\rangle_P + 2 p_{\rm WAIC}\,\quad p_{\rm WAIC} = 2\ln\langle\mathcal{L}\rangle_P - 2\langle\ln\mathcal{L}\rangle_P \,.
\end{equation}
An alternative definition for the model complexities $p_{\rm DIC}$ and $p_{\rm WAIC}$ is based on the variance of the log-likelihood \citep{Watanabe2010}: $p_{\rm DIC} = 2 p_{\rm WAIC} = 2\Var_P\bb{\ln\mathcal{L}}$, which corresponds to the Bayesian model dimensionality used in \citet{Handley2019}. 
We found this definition to be less stable, however, as in certain cases it predicted model dimensionalities larger than the number of varied parameters. 
The stability can be improved in the case where the analysis uses many independent data \citep{Gelman2014,Vehtari2017} but this does not apply to the present case, where we only have access to $\mathcal{O}(1)$ data.
For this reason we use the definitions in Eqs.~\eqref{equ:dicdef} and \eqref{equ:waicdef}.

The final model selection criterion is the Bayes ratio, the ratio of the evidences of the two models under consideration, where the evidence is defined as
\begin{equation}
    \label{equ:evidence}
    Z = \int \mathrm{d}^n\vec\theta \mathcal{L}(\vec\theta)\pi(\vec\theta)\,,
\end{equation}
the integral of the likelihood times the prior $\pi(\vec\theta) = P(\vec\theta|M)$.

To aid interpretability and comparability of these model selection criteria, we put them on a probability scale: each model in the set of models we want to choose from is assigned a weight between 0 and 1, with the weights in the set normalised to 1.  
These weights can then be interpreted as model probabilities. 
For the DIC and WAIC, we do so analogously to Akaike weights \citep{Akaike1978, McElreath2015, Yao2018}. 
The weight for each of the $N$ models under consideration is
\begin{equation}
\label{equ:aicweight}
    w_i = \frac{\mathrm{e}^{-\frac{1}{2}\Delta_i}}{\sum_{j=1}^N \mathrm{e}^{-\frac{1}{2}\Delta_j}} \,,
\end{equation}
where $\Delta_i$ is the difference in DIC (WAIC) between model $i$ and the model with the lowest DIC (WAIC).
The evidences $Z_i$ are already probabilities, such that we only need to normalise them as
\begin{equation}
\label{equ:zweight}
    w_i = \frac{Z_i}{\sum_{j=1}^N Z_j} \,.
\end{equation}
Unless otherwise specified, the sets of models consist of two members: the fiducial, flat $\Lambda$CDM model, and the alternative model under consideration. 

Evaluation of the model selection criteria is subject to uncertainties in the sampling and optimisation procedures. 
We use nested sampling to estimate our posteriors and evidences, where the prior volume of the likelihood contours associated with each sample is not known exactly but only probabilistically \citep{Skilling2006}.  
We follow \citet{Handley2019} and generate many realisations of the prior volumes using \software{anesthetic}\footnote{\url{https://github.com/williamjameshandley/anesthetic}} \citep{anesthetic} to estimate the uncertainties on our DIC, WAIC, and evidence estimates inherent to the sampling procedure. 
Other quantities estimated from nested sampling, such as parameter constraints, are in principle also subject to these uncertainties in the prior volumes. 
We find these uncertainties to be negligible for our parameter constraints, however. 
For example, in the case of $S_8$, this sampling uncertainty is of the order of 1\% of the parameter uncertainty.
We estimate the uncertainty of the value for $\ln \mathcal{L}(\vec\theta_\mathrm{MAP})$ from the scatter of 18 optimisation runs with different starting points. 

\subsection{Tension metrics}
There has been a persistent trend of weak lensing analyses finding lower values of $S_8$ than \textit{Planck}, at varying level of significance \citep[e.g.][]{Heymans2013,MacCrann2015,jee/etal:2016,Joudaki2017, Troxel2018, HSC-Cl, HSC-xi,Joudaki2020,Asgari2020}, with many finding $S_8$ values that are formally consistent with \textit{Planck}, but none finding values higher than \citet{Planck2020-Cosmology}. 
Assessing the agreement or disagreement between data sets is thus a key part of this analysis. 
Here we follow \citet{Heymans2021} in quantifying the concordance or discordance between our results and the temperature and polarisation data from \textit{Planck}. 

We consider three tension metrics to quantify the agreement in a single parameter. 
While all of them agree in the case of Gaussian posterior distributions, their exact values differ when departing from Gaussianity. 
In case of differences between the metrics, we quote the range spanned by them. 
The first compares the distance between the means in the parameter $\theta$ of two data sets A and B to their variances: 
\begin{equation}
\label{equ:tdef}
T(\theta) = \frac{|\overline{\theta}^{\rm A}-\overline{\theta}^{\rm B}|}{\sqrt{\mathrm{Var}[\theta^{\rm A}] + \mathrm{Var}[\theta^{\rm B}]}} \,.
\end{equation}
This metric is exact in the case of Gaussian posteriors. 
To address the cases where the posteriors under consideration depart from Gaussianity, we also consider the Hellinger distance 
\begin{equation}
\label{equ:hellinger}
    d_{\rm H}^2 \bb{ p; q} = \frac{1}{2} \int {\rm d}\theta \bb{ \sqrt{p(\theta)} - \sqrt{q(\theta)} }^2\,,
\end{equation}
where $p(\theta)$ and $q(\theta)$ are the marginal posterior distributions under consideration. 
Finally, we also check the distribution of the parameter shifts, and its associated tension measure
\begin{equation}
    \label{equ:ps}
	p_{\rm S}(\vec\theta) = \int_{P(\Delta\vec\theta)>P(0)} P(\Delta\vec\theta)\diff\Delta\vec\theta \, ,
\end{equation}
where $P(\Delta\vec\theta)$ is the distribution of $\Delta\vec\theta=\vec\theta^{\rm A}-\vec\theta^{\rm B}$.
We refer the reader to appendix G in \citet{Heymans2021} for details. 

Where we want to assess the agreement or disagreement over the whole model, rather than specific parameters, we use the Bayes ratio between a model that jointly describes two data sets and a model that has separate parameters for each of the data sets. 
The Bayes ratio is, however, dependent on the prior choices. 
The suspiciousness \citep{Handley2019} approximately cancels this prior dependence by subtracting the Kullback-Leibler divergence between the posterior and prior. 
As a result, the suspiciousness $\ln S$ can be expressed solely in terms of the expectation values of the log-likelihoods \citep{Heymans2021}:
\begin{splitequation}
	\ln S &= \langle\ln\mathcal{L}_{\rm A+B}\rangle_{P_{\rm A+B}} - \langle\ln\mathcal{L}_{\rm A}\rangle_{P_{\rm A}} - \langle\ln\mathcal{L}_{\rm B}\rangle_{P_{\rm B}} \,.
\end{splitequation}
Finally, we also quote the $Q_{\rm DMAP}$ statistics \citep{Raveri2019}, which measures the change in the best-fit $\chi^2$ values when combining data sets.

\begin{table}
\caption{Sampled parameters and priors.}              
\label{tab:priors}      
\centering                                      
\begin{tabular}{lll}          
\toprule
Parameter & Symbol & Prior \\    
\midrule                                   
Hubble constant & $h$ & $\bb{0.64,\,0.82}$ \\
Baryon density & $\omega_{\rm b}$ & $\bb{0.019,\,0.026}$ \\
CDM density & $\omega_{\rm c}$ & $\bb{0.051,\,0.255}$ \\
Density fluctuation amp. & $S_8$ & $\bb{0.1,\,1.3}$ \\
Scalar spectral index & $n_{\rm s}$ & $\bb{0.84,\,1.1}$ \\
\midrule
Linear galaxy bias (2) & $b_1$ & $\bb{0.5,\,9}$ \\
Quadratic galaxy bias (2) & $b_2$ & $\bb{-4,\,8}$ \\
Non-local galaxy bias (2) & ${\gamma_3^-}$ & $\bb{-8,\,8}$ \\
Virial velocity parameter (2) & $a_{\rm vir}$ & $\bb{0,\,12}$ \\
Intrinsic alignment amp. & $A_{\rm IA}$ & $\bb{-6,\,6}$ \\
Baryon feedback amp. & $A_{\rm bary}$ & $\bb{2,\,3.13}$ \\
Redshift offsets (5) & ${\bf \delta_z}$ & ${\cal N}(\vec{\mu};C_{\delta z})$ \\
SNe absolute calibration & $M$ & $\bb{-22,\,-18}$ \\
\midrule
Curvature & $\Omega_K$ & $\bb{-0.4,\,0.4}$ \\
Sum of masses of neutrinos & $\sum m_\nu$ & $\bb{0,\, 3.0}\,\mathrm{eV}$ \\
Dark energy e.o.s parameter & $w$ & $\bb{-3,\, -0.33}$ \\
$f(R)$-gravity parameter & $\log_{10}|f_{R0}|$ & $\bb{-8,\, -2}$ \\
AGN feedback strength & $\log_{10}{\left(\frac{T_{\rm AGN}}{\mathrm{K}}\right)}$ & $\bb{7.3,\, 8.3}$ \\
\bottomrule
\end{tabular}
\tablefoot{Uniform priors are denoted with square brackets.
The first section lists the primary cosmological parameters, while the second section lists the astrophysical and observational nuisance parameters to model galaxy bias, intrinsic galaxy alignments, baryon feedback, uncertainties in the redshift calibration, and the absolute calibration of SNe. 
The number of separate parameters for each redshift bin is indicated in parentheses. 
The redshift offset parameters are drawn from a multivariate Gaussian prior with mean $\vec{\mu}$ and covariance $C_{\delta z}$. 
The last section lists the priors for the extended parameterisations considered in this work, only one of which is varied at a time. 
Not all parameters are sampled in all analyses. For example, cosmic shear-only results do not vary the galaxy bias parameters.
}
\end{table}

\section{Models}
\label{sec:models}

Here we briefly review the theory behind the $\Lambda$CDM extensions investigated in this work, provide arguments that motivate their analysis, and report recent bounds on their parameters. 

\subsection{Curvature}
The most general line element consistent with translational and rotational symmetries (that is, homogeneity and isotropy) reads 
\begin{equation}
    {\rm d} s^{2}=- c^2 {\rm d} t^{2}+a^{2} \left[{\rm d} \chi^{2}+f_{K}^{2}(\chi) {\rm d} \Omega^{2}\right],
\end{equation}
where $c$ is the speed of light, $\Omega$ denotes the solid angle, $a$ is the scale factor at the cosmic time $t$, $\chi$ is the comoving radial coordinate, and
\begin{equation}\label{eq:com_ang_dist}
f_{K}(\chi)=\left\{\begin{array}{ll}
K^{-1 / 2} \sin \left(K^{1 / 2} \chi\right) & \text { for } K>0 \\
\chi & \text { for } K=0 \\
(-K)^{-1 / 2} \sinh \left[(-K)^{1 / 2} \chi\right] & \text { for } K<0
\end{array}\right.
\end{equation}
is the comoving angular diameter distance, with spatial curvature $K=0$, $K>0$, and $K<0$ producing a flat, closed and open geometry, respectively. The background expansion at late times, ignoring radiation terms, then takes the form 
\begin{equation}\label{eq:expansion_rate_curv}
    \left( \frac{H}{H_0} \right)^2 = \Omega_{\rm m} a^{-3}+\left(1-\Omega_{\rm m}-\Omega_K\right)+\Omega_K a^{-2}\,,
\end{equation}
where $H=\dot a/a$, with the spatial curvature parameter defined as $\Omega_K \equiv -(c/H_0)^2 K$. 
The combination of \textit{Planck} and BAO data provides the tightest constraints to date on this parameter, $\Omega_K=-0.0001\pm0.0018$ at 68\% confidence level \citep{eBOSS2020}, while eBOSS BAO data by themselves constrains curvature to $\Omega_K=0.078^{+0.086}_{-0.099}$. 
However, \textit{Planck} data alone show at least a $3\,\sigma$ preference for a closed universe, with $\Omega_K=-0.044^{+0.018}_{-0.015}$ \citep[68\% CL, but with non-Gaussian tails; ][]{Planck2020-Cosmology}.  

The linear power spectrum is computed with \software{CAMB}, which uses a form of the primordial power spectrum that allows for both curvature and a tilt ($n_{\rm s}\neq1$). 
While this is a phenomenological model, it is commonly used in cosmological analyses, such as \citet{Planck2020-Cosmology}. 
Furthermore, we assume that the non-linear growth of structure in a curved universe can be directly inferred from knowledge of the linear power spectrum alone \citep[cf.][]{Mead2017}, which allows us to use the standard \software{hmcode} prescription \citep{Mead2016}.

\subsection{Massive neutrinos}
The observed neutrino flavour oscillations require at least two of the three neutrino eigenstates $\{ m_1, m_2, m_3 \}$ to be massive \citep{Pontecorvo1958,Fukuda1998,Ahmad2002}, thus cosmologies with $\sum_{i=1}^{3} m_i > 0$ are well-motivated extensions to the base $\Lambda$CDM model. 
Oscillation experiments measure the mass-squared splitting between the mass eigenstates, which provides a lower bound on the sum of neutrino masses. 
In the normal hierarchy ($m_1 < m_2 < m_3$) $\sum m_\nu \gtrsim 0.06 \, {\rm eV}$, while in the inverted hierarchy ($m_3 < m_1 < m_2$) $\sum m_\nu \gtrsim 0.1 \, {\rm eV}$.
Direct measurements of the beta decay of tritium have constrained the mass of the anti-electron neutrino to $m_{\nu_{\rm e}} < 1.1\,{\rm eV}$\citep{Aker2019} at 90\% CL. 

Contrary to cold dark matter, cosmological neutrinos possess high thermal velocities which prevents them from clustering on scales smaller than their free-streaming length, thus suppressing the growth of structure \citep[see, e.g.][]{Lesgourgues2006}. 
Therefore the large-scale structure is a sensitive probe of the sum of neutrino masses, with current constraints in the range $\sum m_\nu < 0.14\text{--}4.5 \, {\rm eV}$ at 95\% CL depending on the particular data set combination and analysis method employed \citep{Lattanzi2017}. 

In this work we assume the normal hierarchy, although our data are not sensitive to this choice. 
The non-linear matter power spectrum is computed with a version of \software{hmcode} \citep{Mead2016} where we removed the contribution of massive neutrinos from the halo mass in the one-halo term in order to provide a better match of \software{hmcode} to the Mira Titan emulator \citep{Lawrence2017} for high neutrino masses \citep[c.f.,][]{Mead2021}. 
This has a suppressing effect on the highly non-linear portion of the \software{hmcode} prediction that scales with the neutrino fraction, being approximately per-cent level for $\sum m_\nu=0.06\,{\rm eV}$.

\subsection{Dark energy equation of state}
Although the cosmological constant phenomenology is in remarkable agreement with a diverse array of observations, the physical mechanism driving the late-time cosmic acceleration remains unknown. 
The simplest possible phenomenological extension to $\Lambda$ is a smooth evolving dark energy component parametrised by a constant equation of state (e.o.s) parameter $w < -1/3$, which matches the cosmological constant for $w=-1$.  The background expansion in these models is modified as
\begin{equation}
    \label{equ:wcdm-background}
    \left( \frac{H}{H_{0}} \right)^2 = \Omega_{\rm m} a^{-3}+\left(1-\Omega_{\rm m}\right) a^{-3(1+w)} \,.
\end{equation}
In principle, Eq.~\eqref{equ:wcdm-background} can include the curvature terms of Eq.~\eqref{eq:expansion_rate_curv} as well but in this work we only consider the cases of either a non-flat Universe or one with an evolving dark energy component.
In our $w$CDM analysis, we assume a single fluid dark energy model with a constant $w$ and a constant sound-speed of $c_{\rm s}^{2} = 1$ (in natural units). 

Previous $3\times2$pt analyses found $w < -0.73$ at 95\% CL, using the previous KiDS release (KiDS-450) combined with 2dFLenS and BOSS spectroscopy \citep{Joudaki2018}, and $w = -0.82^{+0.21}_{-0.20}$ at 68\% CL from DES Y1 imaging data alone \citep{DES-3x2pt}. 
This can be compared with constraints from \textit{Planck} temperature and polarisation data, where $w = -1.58^{+0.52}_{-0.41}$ \citep[95\% CL;][]{Planck2020-Cosmology}, and eBOSS BAO data, which constrain $w=-0.69\pm0.15$ \citep[68\% CL;][]{eBOSS2020}. 
Joint analyses of of earlier \textit{Planck} data, together with CMB data from the South Pole Telescope and a range of non-CMB data found $w=-0.989\pm0.032$ \citep[68\% CL;][]{Park2020}.
Combining \textit{Planck} temperature and polarisation data, eBOSS BAO data, and the Pantheon SNe sample constrains $w = -1.026\pm 0.033$ \citep[68\% CL;][]{eBOSS2020}. 

\subsection{$f(R)$ gravity}
The standard cosmological model rests on the assumption that Einstein’s general relativity (GR) is the correct theory of gravity. 
Departures from GR are tightly constrained on Solar System and astrophysical scales \citep{Will2014,Abbott2017,Sakstein2020,Desmond2020}, but interesting deviations are still possible on larger scales \citep[see, e.g.][]{Joudaki2018,DES-extended,SpurioMancini2019}. 
A breakdown of GR flagged by the large-scale structure statistics would revolutionise the foundations of physics, and could provide an explanation for the observed cosmic acceleration \citep[see, e.g. ][]{Koyama2018, Ferreira2019}.

In this work we focus on $f(R)$ gravity, a popular extension to GR where the Ricci scalar, $R$, is promoted to a generic non-linear function, $f(R)$. 
More specifically, we adopt the Hu-Sawicki functional form, where the range of the fifth force -- the Compton wavelength -- today is given by \citep{Hu2007}
\begin{equation}
    \lambda_{\mathrm{C} 0} \approx 42 \sqrt{\frac{1}{4-3 \Omega_{\mathrm{m}}} \frac{\left|f_{R 0}\right|}{10^{-4}}} \quad h^{-1} \mathrm{Mpc} \,.
\end{equation}
Here $f_{R0}$ is a parameter controlling the extent of the modification, with GR being recovered for $f_{R0}=0$. 
At the level of linear growth the Compton wavelength, $\lambda_{\mathrm{C} 0}$, acts as a cut-off scale.  
On scales $\gg \lambda_{\mathrm{C} 0}$ structures evolve as in GR, whereas on scales $\ll \lambda_{\mathrm{C} 0}$ the gravitational force is enhanced by 1/3. 
In the non-linear regime the activation of the chameleon screening \citep{Khoury2004} drives gravity to GR for values $|f_{R0}| \lesssim 10^{-5}$ \citep[see, e.g.][]{Schmidt2009}. 
Deviations from the $\Lambda$CDM background expansion are $\mathcal{O}(|f_{R0}|)$ \citep{Hu2007}, and since all the models considered here have $|f_{R0}| \ll 1$ we fix the effective equation of state to $w=-1$. 
Using a combination of CMB measurements (or priors) and large-scale structure data the most recent analyses find that values as large as $|f_{R0}| \approx 10^{-5}$ are still consistent with observations at 95\% CL \citep[e.g.][]{Cataneo2015,Liu2016,Alam2016,Hu2016}. 

We compute the non-linear matter power spectrum in $f(R)$ gravity with {\tt ReACT} \citep{Bose2020}, a public \software{C++} library\footnote{\url{https://github.com/nebblu/ReACT}} based on the reaction method of \cite{Cataneo2019}, which we couple to \software{hmcode}. 
The latter provides the cosmology-dependent reference power spectrum to be corrected by the reaction, therefore properly accounting for modified gravity non-linearities. 

\begin{table*}
    \centering
    \caption{Summary of the model selection criteria considered in this work.}
    \label{tab:modelselection}
\begin{tabular}{lccccccc}
    \toprule
    Probe             & $\Delta\chi^2_{\rm MAP}$  & $\Delta{\rm DIC}$ & $\Delta{\rm WAIC}$ & $\Delta\log Z$ & $w_{\rm DIC}$   & $w_{\rm WAIC}$  & $w_{\rm Z}$   \\
    \midrule

   Fix $A_{\rm s}$ (Sect.~\ref{sec:fixAs})\\
    $\quad$Cosmic shear& $0.05\pm0.05$ & $-0.84\pm0.31$ & $-0.54\pm0.19$ & $0.18\pm0.12$ & 0.60  & 0.57  & 0.55  \\
    $\quad$$3\times2$pt& $0.32\pm0.36$ & $-0.71\pm0.71$ & $-0.59\pm0.48$ & $1.66\pm0.27$ & 0.59  & 0.57  & 0.84  \\
[0.3em]   $o\Lambda$CDM (Sect.~\ref{sec:oCDM})\\
    $\quad$Cosmic shear& $-1.25\pm0.08$ & $0.72\pm0.25$ & $-0.00\pm0.17$ & $-0.07\pm0.10$ & 0.41  & 0.50  & 0.48  \\
    $\quad$Galaxy clustering& $0.23\pm0.24$ & $3.92\pm0.55$ & $3.24\pm0.38$ & $-1.10\pm0.26$ & 0.13  & 0.17  & 0.25  \\
    $\quad$$3\times2$pt& $0.10\pm0.34$ & $1.24\pm0.62$ & $0.62\pm0.38$ & $-1.33\pm0.24$ & 0.35  & 0.42  & 0.21  \\
[0.3em]   $\nu\Lambda$CDM (Sect.~\ref{sec:nuCDM})\\
    $\quad$Cosmic shear& $-1.32\pm0.06$ & $-0.27\pm0.25$ & $-0.59\pm0.16$ & $0.29\pm0.11$ & 0.53  & 0.57  & 0.57  \\
    $\quad$Galaxy clustering& $-0.03\pm0.29$ & $2.38\pm0.52$ & $1.77\pm0.34$ & $0.23\pm0.26$ & 0.24  & 0.29  & 0.56  \\
    $\quad$$3\times2$pt& $-0.96\pm0.47$ & $1.59\pm0.70$ & $0.38\pm0.39$ & $0.40\pm0.22$ & 0.31  & 0.45  & 0.60  \\
[0.3em]   $w$CDM  (Sect.~\ref{sec:wcdmres})\\
    $\quad$Cosmic shear& $-1.58\pm0.13$ & $2.43\pm0.26$ & $0.92\pm0.16$ & $-0.38\pm0.11$ & 0.23  & 0.39  & 0.41  \\
    $\quad$Galaxy clustering& $-0.20\pm0.31$ & $4.75\pm0.59$ & $3.24\pm0.37$ & $-0.75\pm0.29$ & 0.09  & 0.17  & 0.32  \\
    $\quad$$3\times2$pt& $0.34\pm0.37$ & $1.53\pm0.61$ & $1.28\pm0.40$ & $-1.85\pm0.25$ & 0.32  & 0.35  & 0.14  \\
[0.3em]   $f(R)\Lambda$CDM  (Sect.~\ref{sec:modgrav})\\
    $\quad$Cosmic shear& $-0.56\pm0.07$ & $0.58\pm0.28$ & $0.09\pm0.18$ & $-0.21\pm0.13$ & 0.43  & 0.49  & 0.45  \\
[0.3em]   Baryon model (App.~\ref{app:HMCODE})\\
    $\quad$Cosmic shear& $0.32\pm0.07$ & $-0.54\pm0.31$ & $-0.05\pm0.19$ & $0.21\pm0.14$ & 0.57  & 0.51  & 0.55  \\
    $\quad$$3\times2$pt& $0.52\pm0.32$ & $-1.50\pm0.69$ & $-1.01\pm0.46$ & $0.20\pm0.27$ & 0.68  & 0.62  & 0.55  \\

    \bottomrule
\end{tabular}
\tablefoot{The first column lists the probes and models under consideration in this work.
The second column list the change in the $\chi^2$ value at the maximum of the posterior compared to the fiducial results of \citet{Asgari2021-CS} and \citet{Heymans2021}. 
The quoted uncertainty is the scatter between optimisation runs. 
Columns $3-5$ list the three model selection criteria considered in this work: DIC (Eq.~\ref{equ:dicdef}), WAIC (Eq.~\ref{equ:waicdef}), and change in the evidence (Eq.~\ref{equ:evidence}), with the uncertainties due to the stochasticity of nested sampling estimates.
The last three columns list the model probabilities based on the three selection criteria with respect to the fiducial, flat $\Lambda$CDM model, as defined in Sect.~\ref{sec:moddelselection}. 
The relative uncertainty on the model probabilities are of the order of 10--20\% but for clarity we do not quote them here.}
\end{table*}

\section{Results}
\label{sec:results}

We first explore how restricting the KiDS-1000 posterior space, either by fixing a subset of parameters to \textit{Planck} best-fit values (Sect.~\ref{sec:fixAs}) or jointly analysing both KiDS and \textit{Planck} with external data sets (Sect.~\ref{sec:extdata}), affects the parameter constraints of KiDS-1000 and their agreement with \textit{Planck}. 
We then explore the effect of extending the parameter space by allowing for curvature ($o\Lambda$CDM, Sect.~\ref{sec:oCDM}), varying the mass of the neutrinos ($\nu\Lambda$CDM, Sect.~\ref{sec:nuCDM}), varying the dark energy equation of state ($w$CDM, Sect.~\ref{sec:wcdmres}), or considering $f(R)$-gravity (Sect.~\ref{sec:modgrav}) has on the KiDS-1000 parameter constraints and whether these extended models can solve the observed tension of KiDS-1000 with \textit{Planck}.

Unless noted otherwise, parameter constraints are reported as the mode of the joint posterior (MAP), together with the projected joint highest posterior density \citep[PJ-HPD, for details see][]{Joachimi2021} credible intervals. 
The model selection criteria and the $S_8$ tension metrics for \textit{Planck} are summarised in Table~\ref{tab:modelselection} and Table~\ref{tab:tension}, respectively. 
In the case where the numerical values of tension metrics differ, we quote the range spanned by them as a robust estimate of the tension in the presence of non-Gaussian posteriors.

\subsection{Fixing the primordial matter power spectrum}
\label{sec:fixAs}
The KiDS-1000 cosmic shear and $3\times2$pt analyses \citep{Asgari2021-CS,Heymans2021} found the amplitude of the measured signal, chiefly dependent on $S_8$, to be low by about $3\,\sigma$ compared to the value derived from the CMB by \textit{Planck}. 
The parameter $S_8 = \sigma_8\sqrt{\Omega_{\rm m}/0.3}$ is well suited to summarise weak lensing results but its mapping to the parameters used to parameterise CMB anisotropies is complicated. 
It is thus not clear whether the observed differences in $S_8$ are due to differences in the amplitude of the matter power spectrum at early and late times, described by $A_{\rm s}$, or other parameters that affect $S_8$.
To answer this question, we test whether fixing $A_{\rm s}$, the amplitude of the primordial matter power spectrum, to the \textit{Planck} best-fit value ameliorates the observed tension in $S_8$ when analysing the KiDS-1000 cosmic shear and $3\times2$pt data. 

\begin{figure}
	\begin{center}
		\includegraphics[width=\columnwidth]{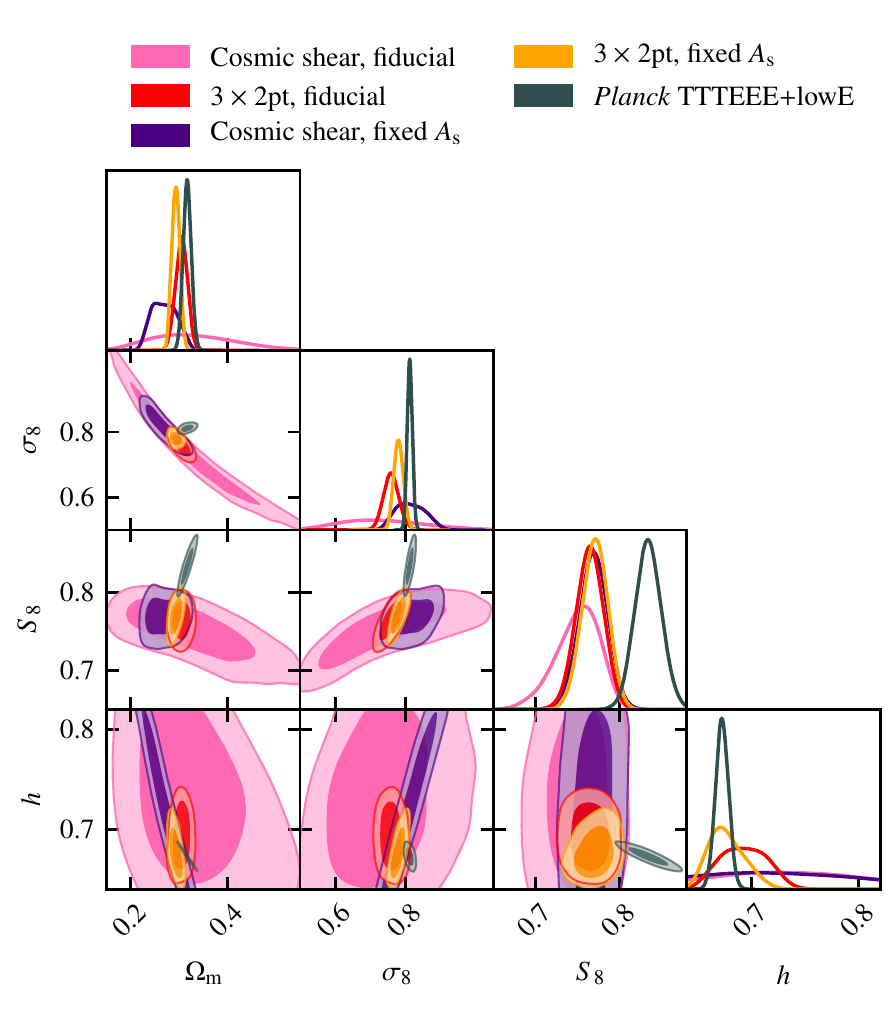}
		\caption{KiDS-1000 cosmic shear and $3\times2$pt parameter constraints when keeping the amplitude of the primordial power spectrum $A_{\rm s}$ fixed to the \textit{Planck} best-fit value. 
		The pink (cosmic shear) and red ($3\times2$pt) contours are the fiducial setup, while the purple (cosmic shear) and orange (cosmic shear) contours show the constraints when $A_{\rm s}$ is being kept fixed. 
		The grey contours denote the \textit{Planck} TTTEEE+lowE results.
		\label{fig:fix_As_ns}}
	\end{center}
\end{figure}

The resulting constraints are shown in Fig.~\ref{fig:fix_As_ns}. 
We find that fixing $A_{\rm s}$ serves to tighten the cosmic shear constraints along the $\Omega_{\rm m}$--$\sigma_8$ degeneracy but does not significantly change the constraints perpendicular to it. 
This is consistent with the known effect of  $A_{\rm s}$ priors primarily affecting the length of the $\Omega_{\rm m}$--$\sigma_8$ `banana' \citep[e.g.][]{Joudaki2017,Chang2019,Joachimi2021} but not constraints on $S_{8}$. 
Fixing $A_{\rm s}$ to the \textit{Planck} best-fit value moves the marginal $S_8$ posterior for cosmic shear to slightly higher values but also reduces its width, such that the tension remains at $2.8\text{--}2.9\,\sigma$. 
For the $3\times2$pt data, the $S_8$ constraints remain largely unchanged, with the tension to \textit{Planck} remaining at $2.9\text{--}3.0\,\sigma$. 
Fixing the tilt of the primordial power spectrum, $n_{\rm s}$, to the \textit{Planck} best-fit value on top of fixing $A_{\rm s}$ does not change these results for either cosmic shear or $3\times2$pt.
The changes in goodness-of-fit when fixing $A_{\rm s}$ lie within our uncertainties on how well we can estimate the $\chi^2$ at the MAP. 
The DIC, WAIC, and Bayes ratio do not disfavour a model with fixed $A_{\rm s}$ either (see Table~\ref{tab:modelselection} for details). 

This highlights that the amplitudes of the two-point statistics of the early-time CMB and the late-time large-scale structure probe different aspects of cosmology. 
While a model with fixed $A_{\rm s}$ still retains enough freedom to describe the cosmic shear and galaxy clustering data, it reduces the freedom in the other parameters. 
Notably, while in the fiducial model the Hubble parameter $h$ is largely uncorrelated with $\Omega_{\rm m}$ and $\sigma_8$, fixing $A_{\rm s}$ induces strong correlations of these parameters with $h$, as seen on the bottom row of Fig.~\ref{fig:fix_As_ns} \citep[c.f.,][]{Sanchez2020}. 
Breaking the induced $\Omega_{\rm m}$--$h$ degeneracy by adding independent information on $\Omega_{\rm m}$ that is consistent with \textit{Planck}, for example through the BAO in the $3\times2$pt data, results in pulling the inferred $h$ constraints down to the \textit{Planck} values. 
On the other hand, breaking the $\sigma_8$--$h$ degeneracy by restricting $\sigma_8$ to \textit{Planck} values results in higher $h$ values, inconsistent with \textit{Planck}. 
In the parameter $S_8$, the $\Omega_{\rm m}$--$h$ and $\sigma_8$--$h$ degeneracies cancel out, so that the $S_8$ constraints and tension with \textit{Planck} are largely independent of $h$.

\begin{table}
    \centering
    \caption{Summary of the tension metrics considered in this work. }
    \label{tab:tension}
\begin{tabular}{lccc}
    \toprule
    Probe             & $T(S_8)$   & $H(S_8)$  & $p_{\rm S}(S_8)$   \\
    \midrule

   Fiducial (flat $\Lambda$CDM)\\
    $\quad$Cosmic shear & $2.8\,\sigma$ & $3.1\,\sigma$ & $3.2\,\sigma$\\
    $\quad$Galaxy clustering & $2.1\,\sigma$ & $2.1\,\sigma$ & $2.1\,\sigma$\\
    $\quad$$3\times2$pt & $3.1\,\sigma$ & $3.1\,\sigma$ & $3.1\,\sigma$\\
[0.3em]   Fix $A_{\rm s}$ (Sec.~\ref{sec:fixAs})\\
    $\quad$Cosmic shear & $2.9\,\sigma$ & $2.8\,\sigma$ & $2.9\,\sigma$\\
    $\quad$$3\times2$pt & $2.9\,\sigma$ & $2.9\,\sigma$ & $3.0\,\sigma$\\
[0.3em]   SNe (Sec.~\ref{sec:extdata})\\
    $\quad$Cosmic shear & $3.0\,\sigma$ & $3.0\,\sigma$ & $3.0\,\sigma$\\
    $\quad$$3\times2$pt & $3.1\,\sigma$ & $3.1\,\sigma$ & $3.0\,\sigma$\\
[0.3em]   CMB lensing (Sec.~\ref{sec:extdata})\\
    $\quad$Cosmic shear & $3.0\,\sigma$ & $3.1\,\sigma$ & $3.0\,\sigma$\\
    $\quad$$3\times2$pt & $2.8\,\sigma$ & $2.8\,\sigma$ & $2.8\,\sigma$\\
[0.3em]   $o\Lambda$CDM (Sec.~\ref{sec:oCDM})\\
    $\quad$Cosmic shear & $2.4\,\sigma$ & $2.5\,\sigma$ & $2.6\,\sigma$\\
    $\quad$Galaxy clustering & $2.4\,\sigma$ & $2.4\,\sigma$ & $2.6\,\sigma$\\
    $\quad$$3\times2$pt & $3.3\,\sigma$ & $2.9\,\sigma$ & $3.0\,\sigma$\\
[0.3em]   $\nu\Lambda$CDM (Sec.~\ref{sec:nuCDM})\\
    $\quad$Cosmic shear & $2.8\,\sigma$ & $2.9\,\sigma$ & $2.9\,\sigma$\\
    $\quad$Galaxy clustering & $1.8\,\sigma$ & $1.8\,\sigma$ & $1.8\,\sigma$\\
    $\quad$$3\times2$pt & $3.4\,\sigma$ & $3.4\,\sigma$ & $3.3\,\sigma$\\
[0.3em]   $w$CDM  (Sec.~\ref{sec:wcdmres})\\
    $\quad$Cosmic shear & $1.3\,\sigma$ & $1.3\,\sigma$ & $1.3\,\sigma$\\
    $\quad$Galaxy clustering & $2.1\,\sigma$ & $2.1\,\sigma$ & $2.1\,\sigma$\\
    $\quad$$3\times2$pt & $1.8\,\sigma$ & $1.8\,\sigma$ & $1.7\,\sigma$\\
[0.3em]   Baryon model (App.~\ref{app:HMCODE})\\
    $\quad$Cosmic shear & $2.3\,\sigma$ & $2.4\,\sigma$ & $2.5\,\sigma$\\
    $\quad$$3\times2$pt & $2.9\,\sigma$ & $2.9\,\sigma$ & $2.9\,\sigma$\\
[0.3em]

    \bottomrule
\end{tabular}
\tablefoot{The first column lists the probes and models under consideration in this work.
The last three columns list the tension in $S_8$ with \textit{Planck} TTTEEE+lowE data using the tension metric $T(\theta)$ (Eq.~\ref{equ:tdef}), the Hellinger distance (Eq.~\ref{equ:hellinger}), and the parameter shift distribution (Eq.~\ref{equ:ps}).}
\end{table}

\subsection{External data: SNe and CMB lensing}
\label{sec:extdata}
Current weak lensing surveys cannot by themselves constrain both $\sigma_8$ and $\Omega_{\rm m}$; the two parameters are degenerate with each other, with the width of degeneracy given by the uncertainty on $\sim S_8$, and its length largely set by the priors \citep{Joudaki2017, Joachimi2021}.
Including external data allows us to break this degeneracy. 
In our $3\times2$pt analysis, this is achieved by the inclusion of spectroscopic galaxy clustering data, which primarily provides constraints on $\Omega_{\rm m}$ through the BAO feature. 

Here we explore two different data sets that allow the breaking of the $\sigma_8$--$\Omega_{\rm m}$ degeneracy; supernovae and lensing of the CMB.
Supernovae provide an independent, low-redshift estimate of $\Omega_{\rm m}$, with our prior on $\omega_{\rm c}$ (see Table~\ref{tab:priors}) being informed by the $5\,\sigma$ constraints on $\Omega_{\rm m}$ derived in \citet{Scolnic2018}.
In CMB lensing, light from the CMB is lensed by the intervening structure between $z=0$ and the surface of last scattering, as detected in the CMB temperature and polarisation anisotropies \citep{Lewis2006}. 
CMB lensing is highly complementary to galaxy lensing, as it exhibits a different degeneracy in the $\sigma_8$-$\Omega_{\rm m}$ plane \citep{Planck2020-CMBlensing}.

\begin{figure}
	\begin{center}
		\includegraphics[width=\columnwidth]{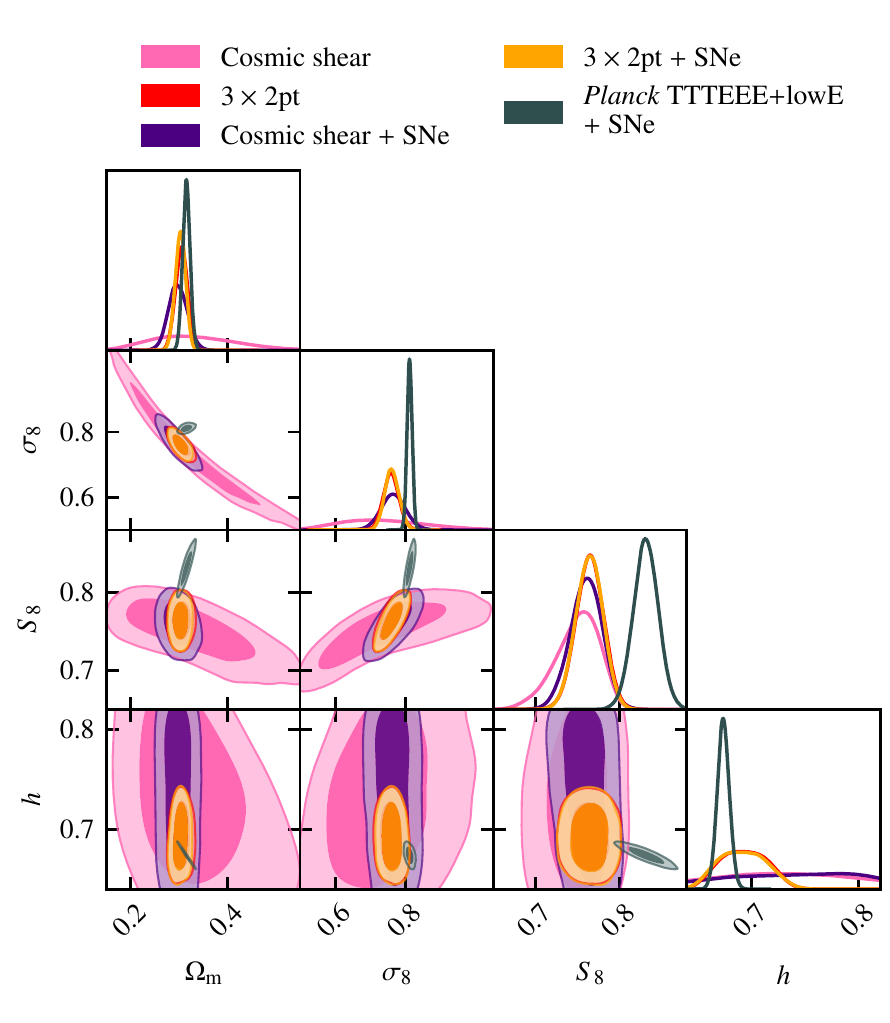}
		\caption{Joint constraints of KiDS-1000 cosmic shear and $3\times2$pt data with the Pantheon supernova data set \citep{Scolnic2018}. 
		The fiducial cosmic shear bandpower and $3\times2$pt results are shown in pink and red, respectively. 
		The joint constraints with Pantheon are denoted in purple and orange, respectively. 
		For $3\times2$pt, the addition of SNe data leaves the constraints virtually unchanged, such that the orange and red contours overlap.
		Finally, the corresponding {\textit{Planck} TTTEEE+lowE + Pantheon} constraints are in grey. 
		\label{fig:sn}}
	\end{center}
\end{figure}

We jointly analyse our cosmic shear bandpower data vector with the Pantheon \citep{Scolnic2018} likelihood, marginalising over the absolute calibration parameter $M$. 
The resulting parameter constraints are summarised in Fig.~\ref{fig:sn}. 
The addition of Pantheon data constrains the matter density to $\Omega_{\rm m}=0.297^{+0.021}_{-0.018}$ and the amount of matter clustering to $\sigma_8=0.769^{+0.028}_{-0.041}$. 
This tightens the constraints on $S_8$ by $\sim 45\,\%$ to $S_8=0.765^{+0.015}_{-0.022}$. 
The increase in constraining power is largely driven by the tight constraints on $\Omega_{\rm m}$ and the residual correlation between $\Omega_{\rm m}$ and $S_8$ in our bandpower cosmic shear results. 
This is made evident by considering the constraints on the parameter $\Sigma_8 = \sigma_8\left(\Omega_{\rm m}/0.3\right)^{0.58}$, which provides a better description of the degeneracy direction in $\Omega_{\rm m}$ and $\sigma_8$ \citep{Asgari2021-CS}: the constraints on $\Sigma_8$ tighten by only $\sim 5\,\%$ when jointly analysing cosmic shear with Pantheon data.

The results of the joint analysis of our $3\times2$pt data with Pantheon do not differ from the fiducial $3\times2$pt analysis. 
The galaxy clustering data already provides stringent constraints on $\Omega_{\rm m}$, such that the addition of the fully consistent, but weaker, constraints on $\Omega_{\rm m}$ from Pantheon does not further improve the constraining power in flat $\Lambda$CDM. 
Similarly, adding the Pantheon likelihood to the \textit{Planck} TTTEEE+lowE likelihood does not appreciably change the \textit{Planck} constraints. 
The tension in $S_8$ thus remains at $3.0\,\sigma$ when analysing both KiDS-1000 and \textit{Planck} jointly with SNe data.
Since the two estimates of $S_8$ are not independent anymore, the tension is to be understood as conditioned on the SNe data.  
Using a prior on $h$ from \citet{Riess2019} based on the local distance ladder does not change the KiDS-1000 cosmic shear or $3\times2$pt results.

\citet{Planck2020-CMBlensing} analysed the reconstructed lensing potential, as inferred from the CMB temperature and polarisation data, which constrains the parameter combination $\sim\sigma_8 \Omega_{\rm m}^{0.25}$. 
This parameter combination is more sensitive to $\sigma_8$ than is the case for $S_8$ and when combined with the galaxy lensing, breaks both degeneracies. 
When jointly analysing our cosmic shear, respectively $3\times2$pt, data with the CMB lensing data\footnote{For technical reasons, we use the \software{cobaya} \citep{Torrado2020} CMB lensing likelihood \url{https://github.com/CobayaSampler/planck_lensing_external}.}, we do so with the KiDS-1000 prior choices \citep{Joachimi2021, Heymans2021}. 
They chiefly differ from those adopted in \citet{Planck2020-CMBlensing} in $h$ and $n_{\rm s}$: the KiDS-1000 prior on $h$ is uniform on the range $[0.64, 0.82]$, approximately encompassing the $5\,\sigma$ ranges of both the CMB constraints from \citet{Planck2020-Cosmology} and the local distance ladder of \citet{Riess2019}, while the CMB lensing analysis of \citet{Planck2020-CMBlensing} adopted a very wide prior\footnote{Sampling in \citet{Planck2020-CMBlensing} was performed with a uniform prior on $\theta_{\rm MC}$, but restricted to the range $H_0\in[0.4, 1.0]$.} of $[0.4, 1.0]$. 
Conversely, the KiDS-1000 prior on $n_{\rm s}$ is uniform on $[0.84, 1.1]$, while \citet{Planck2020-CMBlensing} imposes a tight Gaussian prior of $n_{\rm s}\sim \mathcal{N}(0.96, 0.02)$. 
These different prior choices do not affect the posteriors in the region of parameter space where the galaxy and CMB lensing constraints overlap, but they affect the range of $\Omega_{\rm m}$ values allowed by CMB lensing. 

\begin{figure}
	\begin{center}
		\includegraphics[width=\columnwidth]{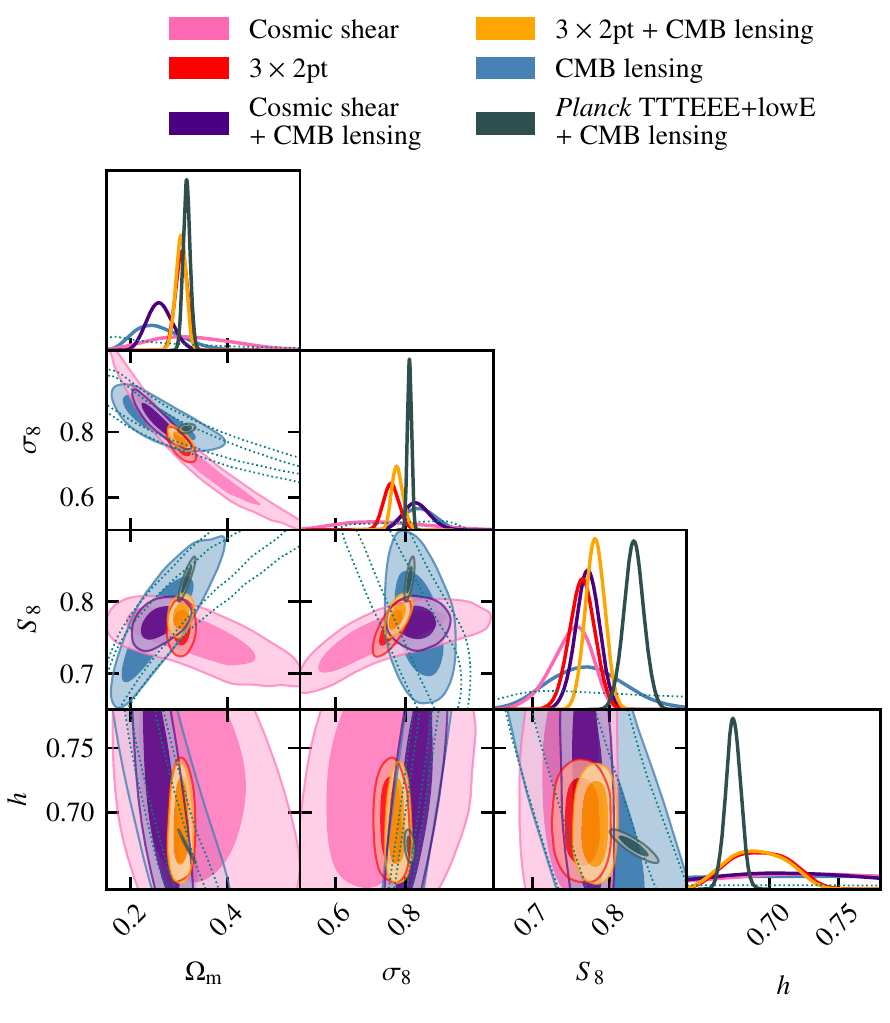}
		\caption{Joint constraints of KiDS-1000 cosmic shear and $3\times2$pt data with CMB lensing data from \citet{Planck2020-CMBlensing}. 
		The fiducial cosmic shear bandpower and $3\times2$pt results are shown in pink and red, respectively, while the joint constraints with CMB lensing are shown in purple and orange, respectively. 
		The \textit{Planck} CMB lensing constraints, with the priors matched to the KiDS analysis, are denoted in solid blue, whereas the fiducial CMB lensing results from \citet{Planck2020-CMBlensing} are denoted with a dotted line. 
		The {\textit{Planck} TTTEEE+lowE} {+ CMB lensing} constraints are shown in grey. 
		\label{fig:cmblensing}}
	\end{center}
\end{figure}

Figure~\ref{fig:cmblensing} illustrates the joint constraints of KiDS-1000 cosmic shear and CMB lensing, as well as KiDS-1000 $3\times2$pt and CMB lensing. 
The combination of KiDS-1000 cosmic shear and CMB lensing constrains the matter density to $\Omega_{\rm m}=0.269^{+0.026}_{-0.029}$, and the clustering amplitude to $\sigma_8=0.81^{+0.047}_{-0.029}$, with $S_8=0.768^{+0.017}_{-0.013}$.
The addition of CMB lensing also improves the $3\times2$pt constraints; we find $\Omega_{\rm m}=0.307^{+0.008}_{-0.013}$, $\sigma_8=0.769^{+0.022}_{-0.010}$, and $S_8=0.779^{+0.013}_{-0.013}$. 
The addition of CMB lensing data thus causes a $\sim 75\,\%$ and $\sim 35\,\%$ improvement in the constraining power on $S_8$ for cosmic shear and $3\times2$pt, respectively. 
As in the case of the joint-analysis with SNe data, the improvement on the cosmic shear $S_8$ constraints is driven by the residual correlation between $\Omega_{\rm m}$ and $S_8$, with the constraints on $\Sigma_8$ tightening by only $\sim 5\,\%$.

Both the cosmic shear and $3\times2$pt-inferred marginal distributions for $S_8$ are narrowed and move to somewhat higher values. 
For cosmic shear, the tension conditioned on the CMB lensing data remains at $3.0\text{--}3.1\,\sigma$, while for $3\times2$pt it is slightly reduced to {$2.8\,\sigma$} (c.f., Table~\ref{tab:tension}).

\subsection{Curvature}
\label{sec:oCDM}

\begin{figure}
	\begin{center}
		\includegraphics[width=\columnwidth]{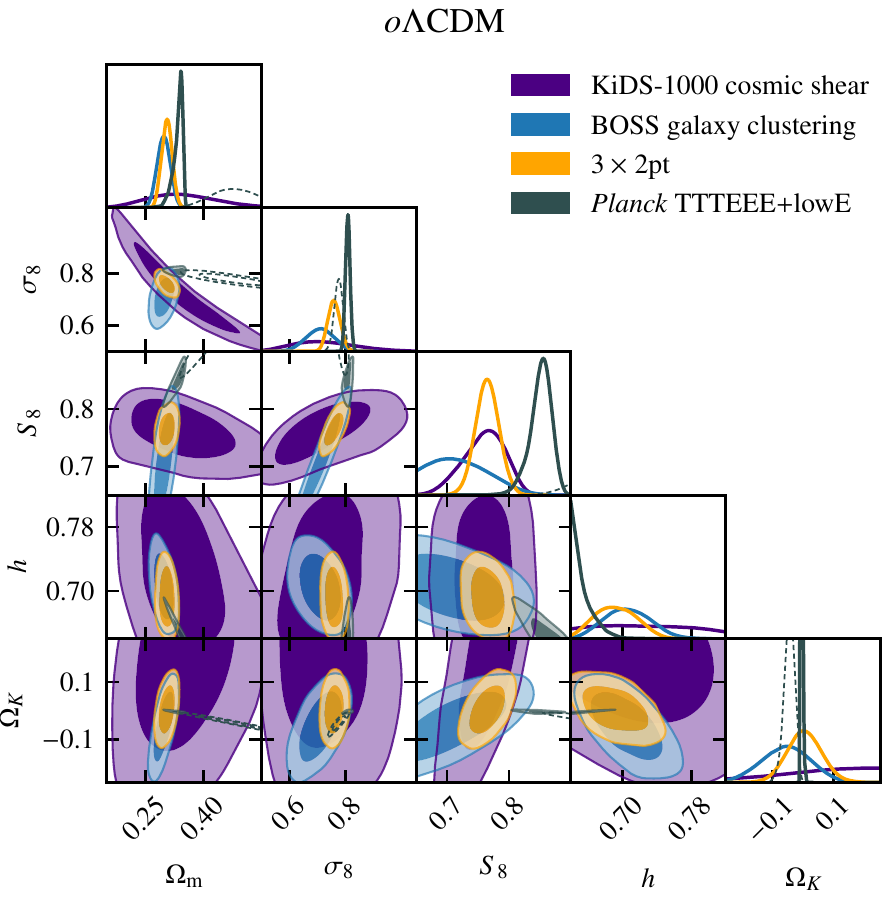}
		\caption{Parameter constraints for a $o\Lambda$CDM model for KiDS-1000 cosmic shear (purple), BOSS DR12 galaxy clustering (blue), and $3\times2$pt (orange). 
		The \textit{Planck} constraints with priors matched to the KiDS setup are shown in solid grey, whereas the fiducial results from \citet{Planck2020-Cosmology} are indicated with a dotted line. 
		\label{fig:kCDM}}
	\end{center}
\end{figure}

We vary $\Omega_K$ uniformly in the interval $[-0.4, 0.4]$, the results of which are shown in Fig.~\ref{fig:kCDM}.  
Our cosmic shear data do not meaningfully constrain $\Omega_K$ but galaxy clustering by itself gives $\Omega_K = -0.07^{+0.12}_{-0.09}$, which is improved on by the full $3\times2$pt data vector to
\begin{equation*}
    \Omega_K = 0.011^{+0.054}_{-0.057}\,. 
\end{equation*}
The \textit{Planck} CMB constraints on $o\Lambda$CDM have significant posterior mass at low values of $h$, outside the KiDS prior range. 
For a comparison to our results, we analyse the \textit{Planck} temperature and polarisation data with the KiDS priors, where we find a disagreement at $2.9\text{--}3.3\,\sigma$ in $S_8$. 
The $o\Lambda$CDM constraints as reported by \citet{Planck2020-Cosmology} prefer a much higher value of $S_8$ due to the preference for high $\Omega_{\rm m}$. 
Compared to these results, the tension is $>4\,\sigma$.
While the priors differ in this case, this has little effect, since our $3\times2$pt results would not change significantly if the $h$ prior were relaxed, as the $S_8$ and $h$ are largely uncorrelated for $3\times2$pt and there is little likelihood mass outside the $h$ prior. 
Our setup of harmonising the priors thus provides a lower bound on the tension in $S_8$. 

The model selection criteria show no preference for the $o\Lambda$CDM model, with it being slightly disfavoured for galaxy clustering and $3\times2$pt but not at any level of meaningful significance.

\subsection{Massive neutrinos}
\label{sec:nuCDM}

\begin{figure}
	\begin{center}
		\includegraphics[width=\columnwidth]{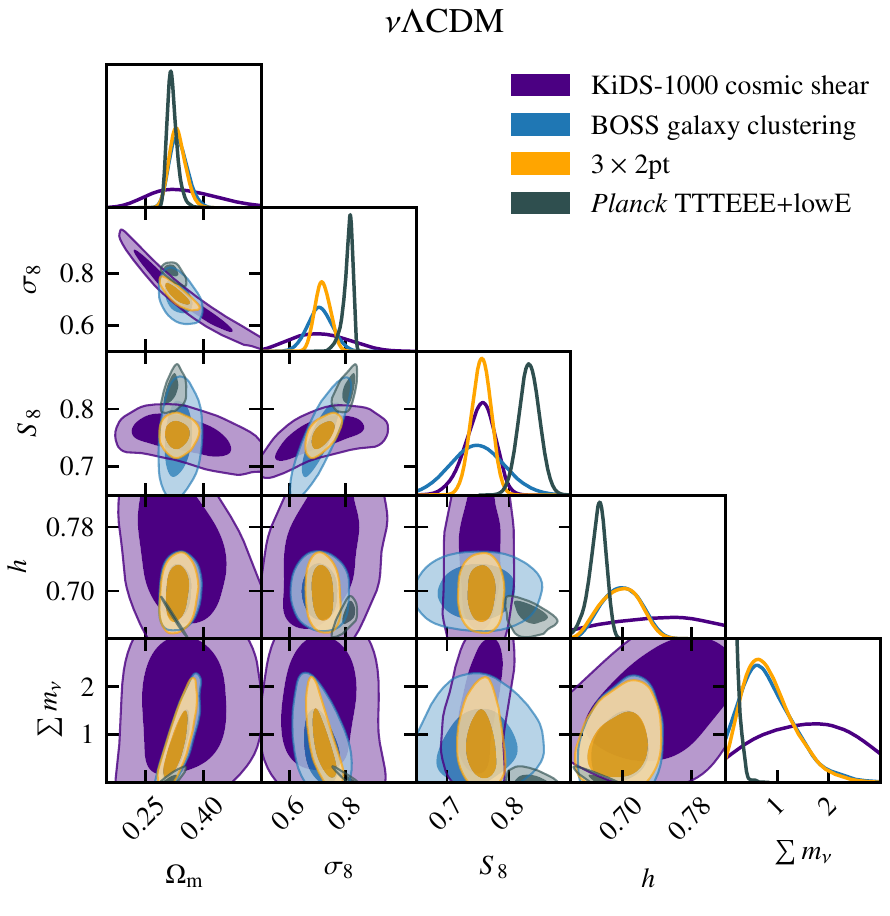}
		\caption{Parameter constraints for a $\nu\Lambda$CDM model for KiDS-1000 cosmic shear (purple) and $3\times2$pt (orange). 
		The \textit{Planck} TTTEEE+lowE constraints are shown in solid grey.
		\label{fig:nuCDM}}
	\end{center}
\end{figure}

The results of varying the sum of the neutrino masses $\sum m_\nu$ uniformly between 0 and $3\,{\rm eV}$ are shown in Fig.~\ref{fig:nuCDM}. 
We find that our $3\times2$pt data provide marginal constraints on the sum of neutrino masses of 
\begin{equation*}
    \sum m_\nu < 1.76\,{\rm eV} \quad \text{(95\% CL)\,.}
\end{equation*}
Allowing the neutrino mass to vary does not affect the cosmic shear constraints but loosens the $3\times2$pt constraints along the cosmic shear $\sigma_8$--$\Omega_{\rm m}$ degeneracy. 
This serves to increase the tension with \textit{Planck} in $S_8$ to $3.3\text{--}3.4\,\sigma$.

Our constraints on $\sum m_\nu$ improve upon earlier results based on KiDS-450, 2dFLenS and BOSS RSD of \citet{Joudaki2018}, who found $\sum m_\nu < 2.2\,{\rm eV}$. 
They also compare favourably to constraints from DES Y1 $3\times2$pt data, when $\sum m_\nu$ was allowed to vary over a larger range\footnote{The constraint is derived from the reanalysis of DES Y1 data in \citet{Planck2020-Cosmology}, available on the Planck Legacy Archive (\url{https://pla.esac.esa.int}).}, which yielded $\sum m_\nu < 2.3\,{\rm eV}$.
They are, however, significantly weaker than other cosmological constraints reported in the literature that include CMB data. 
We believe that combining our constraints with \textit{Planck} in light of the persistent $S_8$ tension would not be a consistent approach, however. 
The joint analysis of \textit{Planck} and DES Y1 data yielded weaker upper limits than just \textit{Planck} data by themselves due to a slight preference of the DES Y1 data for lower clustering amplitudes than \textit{Planck} \citep{DES-3x2pt,Planck2020-Cosmology}. 
As our $3\times2$pt data similarly prefer low clustering amplitudes and do no exclude high neutrino masses, we do not expect a joint analysis with \textit{Planck} to improve upon \textit{Planck}-only constraints on $\sum m_\nu$.

The model selection criteria indicate no preference of a $\nu\Lambda$CDM model over a model where the neutrino mass is fixed to $0.06\,{\rm eV}$.

\subsection{Dark energy equation of state}
\label{sec:wcdmres}
We vary the dark energy equation of state parameter $w$ with a uniform prior of $w \sim \mathcal{U}(-3.0, -0.33)$. 
The upper end of the prior range is chosen such that the cosmic expansion is accelerating. 
To allow comparison with the flat $\Lambda$CDM results, we again keep the priors on the other parameters the same. 
The prior excludes parts of the \textit{Planck} $w$CDM posterior space with high values of $h > 0.82$.  
This region is, however, inconsistent with local measurements \citep{Dhawan2020} and the combined constraints from \textit{Planck} and SNe or BAO \citep{eBOSS2020}. 

\begin{figure}
	\begin{center}
		\includegraphics[width=\columnwidth]{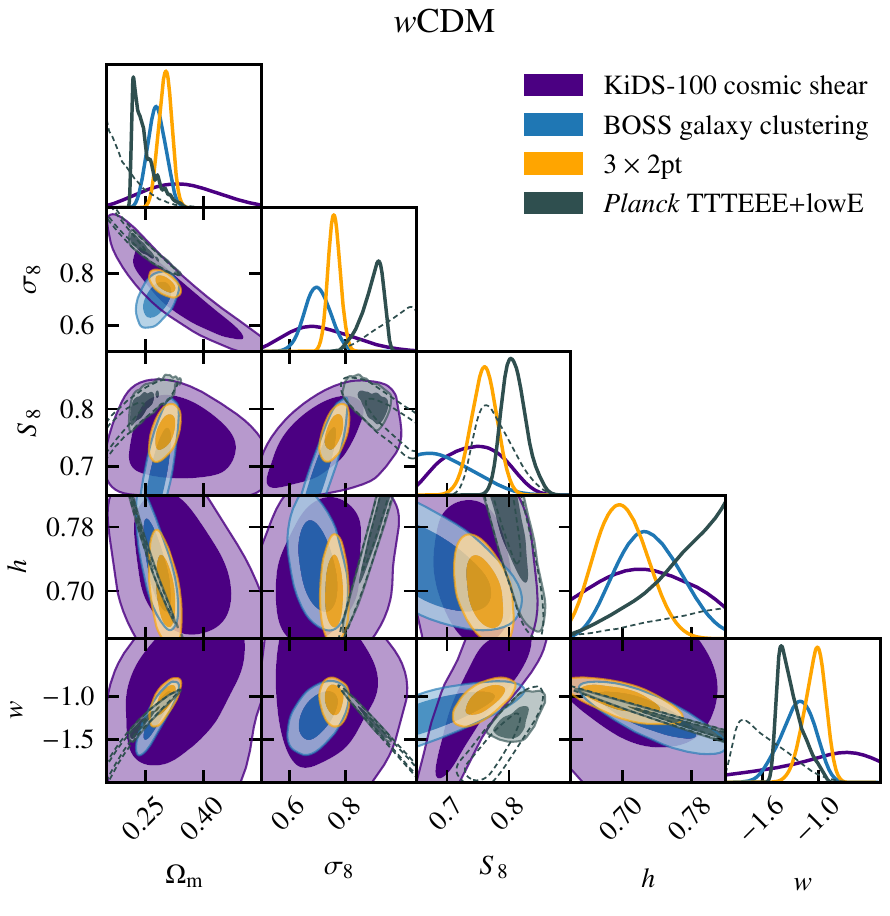}
		\caption{Parameter constraints for a $w$CDM model for KiDS-1000 cosmic shear (purple), BOSS DR12 galaxy clustering (blue), and $3\times2$pt (orange). 
		The \textit{Planck} constraints with priors matched to the KiDS setup are shown in solid grey, while the fiducial results from \citet{Planck2020-Cosmology} are indicated with a dotted line. 
		\label{fig:wCDM}}
	\end{center}
\end{figure}

We present our $w$CDM constraints in Fig.~\ref{fig:wCDM}. 
While our cosmic shear data by themselves do not provide meaningful constraints on $w$, the clustering of the BOSS galaxies does, for which we find $w=-1.05^{+0.21}_{-0.26}$. 
The combination of cosmic shear and galaxy clustering improves the parameter constraints by a factor of about two, with our $3\times2$pt constraints being 
\begin{equation*}
    w=-0.99^{+0.11}_{-0.13}\,. 
\end{equation*}

Among the extensions to the flat $\Lambda$CDM model considered in this work, a $w$CDM model reduces the observed tension on $S_8$ the most, to $1.3\,\sigma$ and  $1.7\text{--}1.8\,\sigma$, respectively for cosmic shear and $3\times2$pt.
The tension in $S_8$ has disappeared due to the marginal \textit{Planck} constraints on this parameter weakening and preferring lower values, especially when allowing for a wide prior in $h$, mirroring previous findings in weak lensing and $3\times2$pt analyses \citep{Joudaki2017-KiDS450-ext, Joudaki2018}. 
We test whether this newfound agreement in $S_8$ extends to other parameters. 
Specifically we assess the agreement in the $S_8$--$w$ parameter space, as well as the agreement on the whole shared parameter space, following the approach in \citet{Heymans2021}. 

To quantify the agreement in the two-dimensional $S_8$--$w$ parameter space, we use the parameter shift statistic Eq.~\eqref{equ:ps}. 
In this space, the tension between our $3\times2$pt constraints and \textit{Planck} is $3.2\,\sigma$. 
Over the full, six-dimensional shared parameter space, there is a $2.1\,\sigma$ tension according the suspiciousness statistic \citep{Handley2019} and a $2.4\,\sigma$ tension according to the $Q_{\rm DMAP}$ statistic \citep{Raveri2019}. 
The Bayes ratio by contrast is $9\pm3$, corresponding to model probabilities of 0.89 vs 0.11 in favour of a single cosmology for both \textit{Planck} and our low-redshift data. 
The Bayes ratio is generally biased towards concordance however, due to essentially double-counting the prior volumes in the case of separate models.
Our model selection criteria do not favour a $w$CDM model but they also do not exclude it at any level of meaningful significance.

\subsection{Modified gravity}
\label{sec:modgrav}
We model the full non-linear effect of $f(R)$ gravity on the matter power spectrum using the reaction formalism \citep{Cataneo2019}. 
The implementation in \software{ReACT} \citep{Bose2020} is currently restricted to modelling the matter power spectrum and does not support modelling of non-linear galaxy bias in modified gravity yet.
We therefore only consider cosmic shear data here. 

\begin{figure}
	\begin{center}
		\includegraphics[width=\columnwidth]{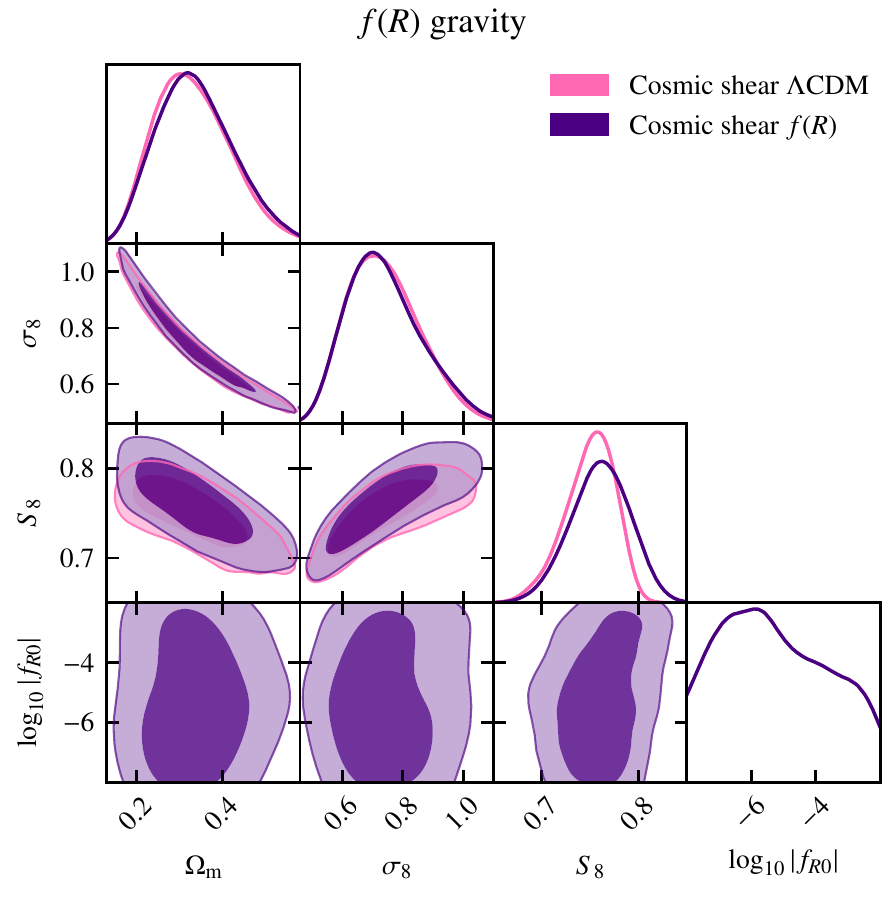}
		\caption{Parameter constraints for a $f(R)$-gravity model for KiDS-1000 cosmic shear (purple), compared to a flat $\Lambda$CDM model (pink). 
		\label{fig:fr}}
	\end{center}
\end{figure}

We sample $\log_{10}|f_{R0}|$ from a uniform prior $\log_{10}|f_{R0}| \sim\mathcal{U}(-8,-2)$ but find that our current cosmic shear data cannot constrain this parameter within this range, as shown in Fig.~\ref{fig:fr}.
While previous work, such as \citet{Harnois-Deraps2015b}, reported constraints of $\log_{10}|f_{R0}| < -4$ from cosmic shear alone, they did not marginalise over cosmological or nuisance parameters.
Future stage IV weak lensing surveys will be able to provide tight constraints on modified gravity models, however, such as $f(R)$ gravity and the DGP \citep{Dvali2000} braneworld models \citep{Bose2020}.

Allowing $f_{R0}$ to vary extends the allowed values of $S_8$ to slightly higher values and could thus in principle serve to reduce the tension with \textit{Planck}.
This is due to the modified gravity linear power spectrum being enhanced in the presence of $f(R)$ gravity, and the derived values of $\sigma_8$ are therefore higher \citep{Planck2016-fR, Wang2020}. 
The same effect moves the \textit{Planck} contours to higher $S_8$ as well, however, such that this is an unlikely mechanism to resolve the observed $S_8$ tension.

\section{Conclusions}
\label{sec:conclusions}
We analysed the KiDS-1000 cosmic shear data and its combination with BOSS and 2dFLenS into a $3\times2$pt data vector in light of extensions to the flat $\Lambda$CDM concordance model of cosmology, external data sets, and restricting the freedom of the model.

We found that restricting the freedom of the model to set the amplitude of the primordial power spectrum through $A_{\rm s}$ does not, maybe surprisingly, resolve the tension with \textit{Planck} in the late-time amplitude parameter $S_8$. 

Jointly analysing our cosmic shear and $3\times2$pt data with external data sets, namely Type Ia SNe and CMB lensing, serves to break parameter degeneracies, improving the KiDS-1000 cosmic shear constraints in $S_8$ by $\sim 45\,\%$ in case of SNe, and $\sim 75\,\%$ in the case of CMB lensing. 
The improvement on the cosmic shear constraints is more modest at $\sim 5\,\%$ when considering the parameter $\Sigma_8 = \sigma_8\left(\Omega_{\rm m}/0.3\right)^{0.58}$, which captures the $\Omega_{\rm m}$-$\sigma_8$ degeneracy better. 
Neither of these external data sets are able to pull the \textit{Planck} and KiDS-1000 constraints on $S_8$ closer together, however. 

Using three model selection criteria, we assessed whether the data prefer a model other than flat $\Lambda$CDM but we found that none of the extensions considered are favoured or disfavoured. 
We provide constraints independent of the CMB on the curvature $\Omega_K=0.011^{+0.054}_{-0.057}$ and dark energy equation of state parameter $w=-0.99^{+0.11}_{-0.13}$, both of which are fully consistent with their flat $\Lambda$CDM values. 
The constraints on $w$ are tighter than those from either eBOSS BAO or \textit{Planck} temperature and polarisation data alone but weaker than their combination. 
Neither of these extensions are preferred by the data over the fiducial flat $\Lambda$CDM model according to a range of model selection criteria.

Our data are only able to provide weak constraints on the sum of the neutrino masses $\sum m_\nu < 1.76\,{\rm eV}$ at 95\% CL. 
They are, however, independent of CMB data. 
We use a full non-linear modelling for the matter power spectrum to constrain $f(R)$ gravity but find that current weak lensing data can not constrain $f_{R0}$ by itself. 
Future weak lensing data, as well as the joint analysis with external data sets will be able to improve these constraints significantly \citep{Bose2020}. 

We find that the $\sim 3\,\sigma$ tension with \textit{Planck} CMB data that was found in \citet{Asgari2021-CS} and \citet{Heymans2021} is not resolved by either extending the parameter space beyond flat $\Lambda$CDM, or by restricting it through fixing the amplitude of the primordial power spectrum to the \textit{Planck} best-fit value.
To further our understanding of this difference between the early and late-time Universe, we look forward with anticipation to the upcoming independent weak lensing analyses from the Dark Energy Survey and Hyper Suprime-Cam Survey.

\begin{acknowledgements}
We thank Antony Lewis for prompting the investigation of the effect of fixing $A_{\rm s}$ and Will Handley for useful discussions.  

The figures in this work were created with \software{matplotlib} \citep{Hunter2007} and \software{getdist}, making use of the 
\software{numpy} \citep{Oliphant2006} and \software{scipy} \citep{Jones2001} software packages. 
\\

This project has received significant funding from the European Union's Horizon 2020 research and innovation programme. 
We thank and acknowledge support from: 
the European Research Council under grant agreement No.~647112 (TT, MA, MCa, CH, CL, and BG), No.~770935 (HHi, AHW, and AD) and No.~693024 (SJ) in addition to the Marie Sk\l{}odowska-Curie grant agreements No.~797794 (TT) and No.~702971 (AM). 
We also acknowledge support from the Max Planck Society and the Alexander von Humboldt Foundation in the framework of the Max Planck-Humboldt Research Award endowed by the Federal Ministry of Education and Research (CH, FK);  
the Swiss National Science Foundation Professorship grant No.~170547 (BB and LL);
the Deutsche Forschungsgemeinschaft Heisenberg grant Hi 1495/5-1 (HHi);  
the Netherlands Organisation for Scientific Research Vici grant 639.043.512 (HHo, AK); 
the Alexander von Humboldt Foundation (KK);  
the Polish Ministry of Science and Higher Education through grant DIR/WK/2018/12, and the Polish National Science Center through grants no. 2018/30/E/ST9/00698 and 2018/31/G/ST9/03388 (MB); 
the Royal Society through an Enhancement Award RGF/EA/181006 (BG);  
the Australian Research Council grants DP160102235 and CE17010013 (KG);  
the Beecroft Trust (SJ);  
the NSFC of China under grant 11973070, the Shanghai Committee of Science and Technology grant No.19ZR1466600, and the Key Research Program of Frontier Sciences, CAS, Grant No. ZDBS-LY-7013 (HYS).\\

Funding for SDSS-III has been provided by the Alfred P. Sloan Foundation, the Participating Institutions, the National Science Foundation, and the U.S. Department of Energy Office of Science. The SDSS-III web site is http://www.sdss3.org/.

SDSS-III is managed by the Astrophysical Research Consortium for the Participating Institutions of the SDSS-III Collaboration including the University of Arizona, the Brazilian Participation Group, Brookhaven National Laboratory, Carnegie Mellon University, University of Florida, the French Participation Group, the German Participation Group, Harvard University, the Instituto de Astrofisica de Canarias, the Michigan State/Notre Dame/JINA Participation Group, Johns Hopkins University, Lawrence Berkeley National Laboratory, Max Planck Institute for Astrophysics, Max Planck Institute for Extraterrestrial Physics, New Mexico State University, New York University, Ohio State University, Pennsylvania State University, University of Portsmouth, Princeton University, the Spanish Participation Group, University of Tokyo, University of Utah, Vanderbilt University, University of Virginia, University of Washington, and Yale University.

Based on data products from observations made with ESO Telescopes at the La Silla Paranal Observatory under programme IDs 177.A-3016, 177.A-3017 and 177.A-3018. 
\\
{ {\it Author contributions:}  All authors contributed to the development and writing of this paper.  The authorship list is given in three groups:  the lead author (TT) followed by two alphabetical groups.  The first alphabetical group includes those who are key contributors to both the scientific analysis and the data products.  The second group covers those who have either made a significant contribution to the data products, or to the scientific analysis.}

\end{acknowledgements}

\bibliographystyle{aa}
\bibliography{references.bib}

\begin{appendix}
\section{Baryonic effects}
\label{app:HMCODE}
\citet{Asgari2021-CS} and \citet{Heymans2021} used the model of \citet{Mead2016}, \software{hmcode}, to predict the non-linear matter power spectrum and marginalise over the effect of baryons. 
The effect of baryons in \software{hmcode} is modelled by a phenomenological `bloating' of the dark-matter halos and changing the halo concentration. 
Recently, \citet{Mead2021} proposed a new, physically motivated modelling approach within the \software{hmcode}-framework, which provides a parameterisation of the effect of feedback from active galactic nuclei (AGN) on the matter power spectrum. 
To test whether this new parameterisation affects our cosmology constraints, we vary the parameter $\log_{10}{\left(\frac{T_{\rm AGN}}{\mathrm{K}}\right)}$ over the range $[7.3, 8.3]$, a conservative choice as it extends well beyond the range $7.6 < \log_{10}{\left(\frac{T_{\rm AGN}}{\mathrm{K}}\right)} < 8.0$ found to reproduce observations in the BAHAMAS suite of hydrodynamical simulations \citep{McCarthy2017}. 
Higher values of $\log_{10}{\left(\frac{T_{\rm AGN}}{\mathrm{K}}\right)}$ correspond to more violent feedback, which expels more gas from halos, thus lowering the power on intermediate scales.

\begin{figure}
	\begin{center}
		\includegraphics[width=\columnwidth]{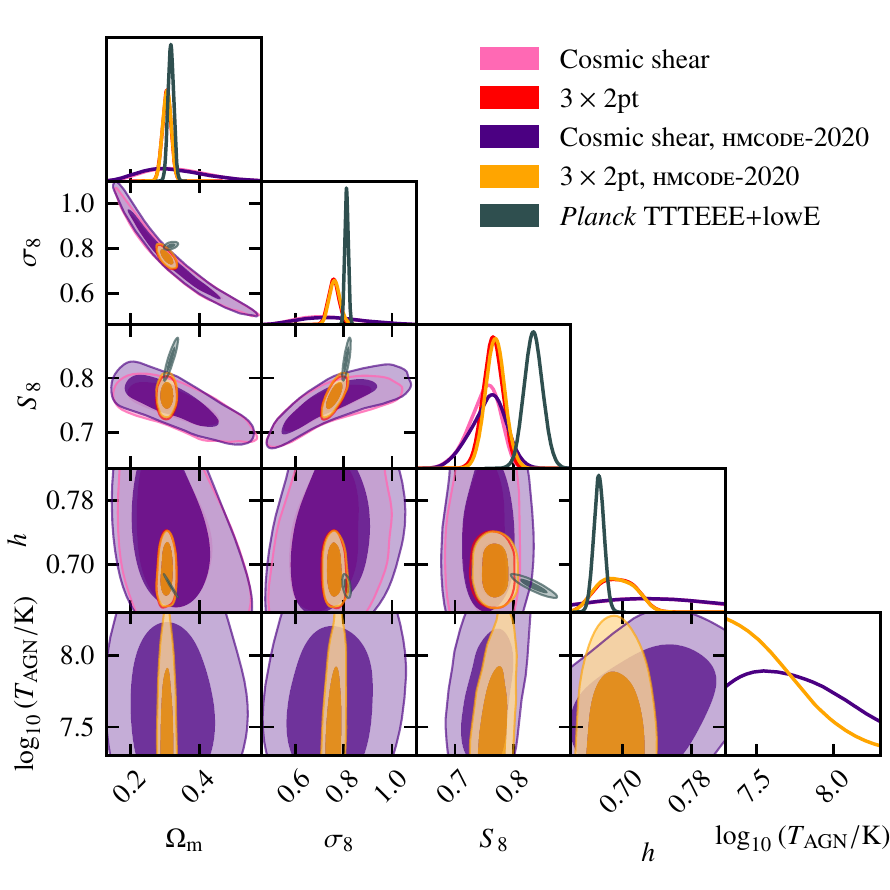}
		\caption{Comparison of the KiDS-1000 cosmic shear and $3\times2$pt parameter constraints for different choices of the non-linear modelling of the matter power spectrum. 
		The pink (cosmic shear) and red ($3\times2$pt) contours are derived using our fiducial setup, using the \citet{Mead2016} \software{hmcode} model.
		The purple (cosmic shear) and orange ($3\times2$pt) contours use the updated \citet{Mead2021} model with a physically motivated modelling of baryonic effects. 
		The \textit{Planck} TTTEEE+lowE contours are shown in grey.
		\label{fig:hmcode}}
	\end{center}
\end{figure}

The resulting parameter constraints are presented in Fig.~\ref{fig:hmcode}. 
We find good agreement with the result based on the previous version of \software{hmcode} \citep{Mead2016}. 
The preference for low values of $\log_{10}{\left(\frac{T_{\rm AGN}}{\mathrm{K}}\right)}$ is consistent with the preference for high values of $A_{\rm bary}$ in the KiDS-1000 cosmic shear and $3\times2$pt data.
We caution against a too literal interpretation of this parameter, as other effects can mimic the suppression of the matter power spectrum at intermediate to small scales.

Unlike the purely phenomenological modelling of the effect of baryonic processes in \software{hmcode}, the model in \software{hmcode-2020} is more physically motivated, including gas and stellar components. 
As such, it does not have a dark matter-only limit, as diffuse gas always causes a degree of suppression of power at intermediate scales and stars cause an increase of power at the smallest scales. 
Furthermore, the suppression of power due to AGN feedback sets in at larger scales, with the strongest feedback scenario considered here, $\log_{10}{\left(\frac{T_{\rm AGN}}{\mathrm{K}}\right)}=8.3$, exhibiting a stronger suppression of the matter power spectrum for $k\lesssim 10\,h{\rm Mpc}^{-1}$ than the strongest feedback scenario considered in the fiducial analysis, $A_{\rm bary}=2$. 
This model-inherent suppression of power serves to exclude low values of $S_8$, while the freedom of the model to predict a strong suppression due to our wide prior on $\log_{10}{\left(\frac{T_{\rm AGN}}{\mathrm{K}}\right)}$ allows for high values of $S_8$.
Together, these effects cause a shift of the marginal $S_8$ posterior to slightly larger values, reducing the tension of our cosmic shear results with \textit{Planck} from $2.8$--$3.2\,\sigma$ in the fiducial case to $2.3$--$2.5\,\sigma$ when using \software{hmcode-2020}. 
This better agreement in $S_8$ is partially driven by the stronger correlation between $\Omega_{\rm m}$ and $S_8$ in the case of the \software{hmcode-2020} model. 
Using $\Sigma_8 = \sigma_8\left(\Omega_{\rm m}/0.3\right)^{0.58}$ instead, which provides a better description of the degeneracy direction in $\Omega_{\rm m}$ and $\sigma_8$ \citep{Asgari2021-CS}, reduces the tension from $3.2$--$3.4\,\sigma$ to $2.9\,\sigma$. 
The effect on the $3\times2$pt results is smaller, reducing the tension from $3.1\,\sigma$ to $2.9\,\sigma$. 
The shift of the $3\times2$pt best-fit value of $S_8$ in terms of the $S_8$ uncertainty when using the \software{hmcode-2020} model is $0.26\,\sigma$, similar to the shift observed when using \software{halofit} instead of \software{hmcode} \citep{Joachimi2021}. 
This result therefore confirms the conclusions of \citet{Joachimi2021}: the uncertainty in the non-linear matter power spectrum prescription is currently one of the dominant systematics in the modelling and cosmology inference for KiDS.

\section{Extended data cuts and prior choices}
\label{app:nscuts}
\citet{Troester2020} and \citet{Heymans2021} found a preference for low values of the spectral index $n_{\rm s}$ inferred from the clustering and $3\times2$pt analyses. 
It was speculated that large-scale systematics in the galaxy clustering measurement \citep[for BOSS DR12, see][]{Ross2017} could be responsible but they argued that the main cosmological results, namely constraints on $S_8$, are not affected. 
Here we explore this preference for low values of $n_{\rm s}$ further by exploring the effect of data cuts that discard the large-scale information in the clustering measurements, as well as the effect of fixing $n_{\rm s}$, on the remaining cosmological parameters.

\begin{figure*}
	\begin{center}
		\includegraphics[width=\textwidth]{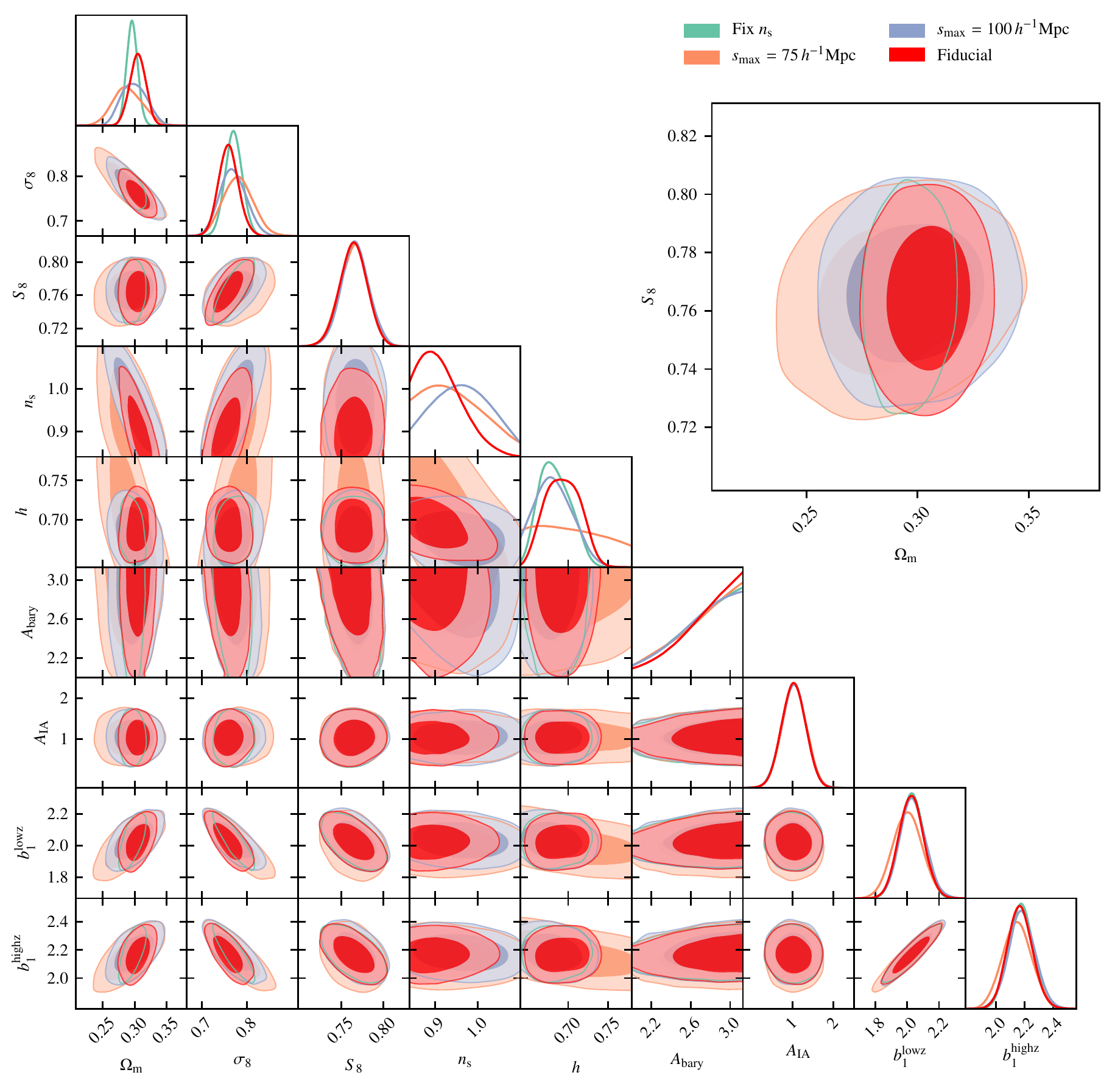}
		\caption{Effect of fixing $n_{\rm s}$ and discarding galaxy clustering data at large scales. 
		Constraints when $n_{\rm s}$ is fixed are shown in turquoise, while those were the maximum separation $s_{\rm max}$ in the correlation function wedges is limited are shown in blue ($s_{\rm max}=100\,h^{-1}{\rm Mpc}$) and orange ($s_{\rm max}=75\,h^{-1}{\rm Mpc}$), compared to the fiducial setup in red. 
		\label{fig:largescalesystematics}}
	\end{center}
\end{figure*}

The resulting constraints are shown in Fig.~\ref{fig:largescalesystematics}. 
Excising large-scale galaxy clustering data from the $3\times2$pt data vector by limiting the maximum separation in the correlation function wedges to $s_{\rm max}=100\,h^{-1}{\rm Mpc}$ or $s_{\rm max}=75\,h^{-1}{\rm Mpc}$ primarily degrades the constraining power in $\Omega_{\rm m}$ as a consequence of removing the information about the BAO peak.
These scale cuts only cause small changes in other parameters and leave $S_8$ unchanged.

In a similar vein, fixing $n_{\rm s}$ breaks its degeneracies with $\Omega_{\rm m}$ and $\sigma_8$, resulting in slightly tighter constraints on these parameters but leaving $S_8$ unaffected. 
We thus conclude that our analysis is robust to these systematics.

\end{appendix}

\end{document}